# Three-dimensional sedimentation patterns of two interacting disks in a viscous fluid

Yi Liu[1], Yu Guo[1, 3, †], Bo Yang[2], Dingyi Pan[1], Zhenhua Xia[1],

Zhaosheng Yu[1], and Lian-Ping Wang[2, †]

[1] Department of Engineering Mechanics, Zhejiang University, Hangzhou 310027, China

[2] Department of Mechanics and Aerospace Engineering, Southern University of Science and Technology, Shenzhen 518055, China

[3] State Key Laboratory of Clean Energy Utilization, Zhejiang University, Hangzhou, 310027, China

**Abstract**

The sedimentation of two spherical solid objects in a viscous fluid has been extensively investigated and well understood. However, a pair of flat disks (in three dimensions) settling in the fluid shows more complex hydrodynamic behaviors. The present work aims to improve understanding of this phenomenon by performing Direct Numerical Simulations (DNS) and physical experiments. The present results show that the sedimentation processes are significantly influenced by disk shape, characterized by a dimensionless moment of inertia $I^*$, and Reynolds number of the leading disk $Re$. For the flatter disks with smaller $I^*$, steady falling with enduring contact transits to periodic swinging with intermittent contacts as $Re$ increases. The disks with larger $I^*$ tend to fall in a Drafting-Kissing-Tumbling (DKT) mode at low $Re$ and to remain separated at high $Re$. Based on $I^*$ and $Re$, a phase diagram is created to classify the two-disk falling into ten distinctive patterns. The planar motion or three-dimensional motion of the disks is determined primarily by $Re$. Turbulent disturbance flows at a high $Re$ contribute to the chaotic three-dimensional rotation of the disks. The chance for the two disks to contact is increased when $I^*$ and $Re$ are reduced.

†Email addresses for correspondence: yguo@zju.edu.cn, wanglp@sustech.edu.cn

# 1. Introduction

Sedimentation of particles in a fluid is ubiquitous in industrial applications and environmental processes, such as industrial waste treatment, coal-water slurry transport, proppant transport in hydraulic cracking, soot particle dispersion, falling of leaves, and settling of sands in a river. Understanding the falling patterns is helpful to predict particle trajectories and final locations of the settling particles. Unfortunately, the falling patterns are rather complex due to fluid-particle and particle-particle interactions. Thus, extensive studies have been performed to reveal underlying mechanism of the particle sedimentation in a fluid.

Fortes et al. (1987) observed two spheres settling in a pattern of Drafting-Kissing-Tumbling (DKT) in their experiments, demonstrating the strong hydrodynamic interactions between the particles. In the two-dimensional (2D) numerical simulations of two circular particles falling in a long narrow channel (Aidun and Ding, 2003), effects of confinement from the channel walls on dynamics of the particles are significant as the width of channel $L_w$ is only 4 times particle diameter $d_p$. The particle motion is driven by a dimensionless weight of a particle, $W^* = \pi G_a(\rho_r - 1)/4$, in which $G_a$ is the Galileo number ($G_a = d_p^3 g/\nu_f^2$, $g$ is the gravitational acceleration, and $\nu_f$ is the kinematic viscosity of the fluid) and $\rho_r$ is the density ratio of particle to fluid. Under the channel boundary confinement of $L_w/d_p = 4$, the falling styles of the particles are controlled by the parameter $W^*$. An increase in $W^*$ from 100 to 200 leads to an increase in particle terminal Reynolds number $Re$ from 2 to 5. In this regime, as $W^*$ increases, the initial periodic state of the horizontal particle motion transits to another periodic branch. Further increase in $W^*$ results in a cascade of period-doubling bifurcations to a chaotic state represented by a low dimensional chaotic attractor (Aidun and Ding, 2003). For 200 < $W^*$ < 400, corresponding to 6 < $Re$ < 10, the pair of particles fall in the DKT pattern (Feng and Joseph, 1995). At the large driving forces, *i.e.*, 400 < $W^*$ < 500, the two particles tend to form a horizontal structure, causing the maximum effective blockage ratio. The falling speeds are reduced and the particle Reynolds numbers return to the range 5 < $Re$ < 7. Then, the DKT vanishes and the settling experiences a quasi-periodic transition to the chaotic state by increasing the driving force (Verjus et al., 2016). To estimate particle-particle contact duration, a scaling law was proposed by Li et al. (2020), considering the effects of particle-fluid density ratio, fluid viscosity, interparticle friction, and adhesive force.

Various factors can affect the sedimentation patterns. After removing rotational degrees of freedom of the particles, the lateral migration still occurred but the divergent particle oscillation disappeared, making the particles fall with a steady oblique or horizontal structure at smaller terminal Reynolds numbers (Zhang et al., 2018). Sedimentation of two unequal particles of different sizes and densities was numerically investigated (Nie et al., 2021; Nie and Lin, 2020). A pattern of horizontal oscillatory motion, characterized by a structure with a large and light particle above a small and heavy one and strong oscillations of both particles in the horizontal direction, was observed for diameter ratio of 0.3 at intermediate Reynolds numbers, and the magnitude of oscillations decreased with an increase in the density ratio (Nie et al., 2021). In three-dimensional (3D) sedimentation of two spheres of different densities in a square tube, the spheres oscillated in the central plane of the tube at high Galileo numbers ($G_a$); While at low $G_a$, the spheres moved to one of the diagonal planes of the tube, reaching a steady or periodic state depending on the density difference between them (Nie and Lin, 2020).

The sedimentation of particles in a fluid exhibits a strong dependence on particle shape. The simulation results by Fornari et al. (2018) showed that the mean settling speed of oblate particles was significantly higher than that of spheres in dilute suspensions at the solid volume fractions 0.5-1 %, due to the formation of columnar-like clusters of the oblate particles. The early experimental observations of a single disk falling in a quiescent viscous fluid by Willmarth et al. (1964) and Field et al. (1997) demonstrated the remarkable effect of particle shape, which was characterized by a dimensionless moment of inertia (the ratio of the moment of inertia of a thin disk about its diameter $I_d$ and a quantity proportional to the moment of inertia of a rigid sphere of fluid about its diameter),

$$I^* = \frac{I_d}{\rho_f d_c^5} = \frac{\pi \rho_s l_c}{64 \rho_f d_c} + \frac{\pi \rho_s l_c^3}{48 \rho_f d_c^3}, \qquad (1.1)$$

in which $\rho_s$ and $\rho_f$ are the densities of the solid disk and fluid, respectively; $l_c$ and $d_c$ are the thickness and diameter, respectively, of the disk. For the very thin disks, the thickness $l_c$ is much smaller than $d_c$, thus, the second term, i.e. $\pi \rho_s l_c^3/(48 \rho_f d_c^3)$, on the right hand of the equal sign in the Eq.(1.1) was neglected in the previous work (Willmarth et al., 1964; Field et al., 1997; Zhong et al., 2011). The falling pattern was also dependent on the terminal Reynolds number of the disk *Re*, which was calculated from the mean vertical velocity of the disk during the sedimentation (excluding the initial acceleration stage). Based on $I^*$ and *Re*, Field et al. (1997) classified the falling processes of a single disk into four distinct styles: steady falling, periodic, chaotic, and tumbling. Steady falling pattern occurred at low *Re*. At high *Re*, a disk with a smaller dimensionless moment of inertia $I^*$ settled periodically; as $I^*$ increased, the falling pattern transited from the periodic to the chaotic and then to the tumbling. With more experimental observations, Zhong et al. (2011) partitioned the periodic falling pattern further into three sub-modes: planar zigzag, transitional, and spiral. It was believed that the transition from the 2D zigzag motion to the 3D spiral motion occurred due to the growth of 3D disturbances (Lee et al., 2013). In their numerical simulations, Auguste et al. (2013) observed two additional 3D sedimentation patterns in which the disk experienced a slow horizontal precession superimposed on zigzagging or tumbling motions. For the infinitely thin disks with very small $I^*$, Chrust et al. (2013) created a comprehensive phase diagram of transition scenarios of the different settling types, based on the Galileo number and a dimensional mass. The hydrodynamics and typical nonlinear paths of various non-spherical bodies freely falling or rising in fluids were systematically reviewed by Ern et al. (2012).

The interaction of two identical disks falling in tandem in a fluid was experimentally investigated by Brosse and Ern (2011; 2014). It was observed that the trailing body would catch up with the leading one for vertical separation distances up to 14 diameters under the condition of a horizontal separation distance less than 0.5 diameters. The wake of the leading body had an impact on the fluid drag force exerted on the trailing one, depending on the distance between them. Thick disks separated laterally after the collision and eventually fell side by side. At low *Re* of 115 and 152, thinner disks formed a steady Y-configuration by contacting each other and fell together in rectilinear paths, as the wakes of the two bodies merged, creating strong interbody attraction. Such steady falling with a small relative inclination between the two bodies was also obtained in the numerical simulations of the sedimentation of two oblate ellipsoids (Ardekani et al., 2016). At higher *Re* of 255 and 275, the thinner disks exhibited inclined periodic motions with the relative distance and inclination of the bodies fluctuating in time. The periodic vortex shedding played a role in the oscillatory motions of the two disks. In addition, the inhomogeneity of the wake of the leading body destabilized the wake of the trailing one, amplifying the oscillation of the trailing one. In the experiments of two identical disks falling side by side in a fluid at rest for *Re* ranging from 100 to 300 (Ern and Brosse, 2014), the two disks repelled each other at the smaller horizontal separation distances and they appeared to move independently at the larger horizontal distances. In the case of oscillatory paths, no synchronization was observed between the paths and between the wakes of the two disks. A repulsion coefficient, which is proportional to the magnitude of repulsion force, was found to decrease with the horizontal separation distance and increase with the thickness-to-diameter ratio of the disks.

Compared to the sedimentation of a pair of circular particles (2D) and spheres (3D), the hydrodynamic behaviors of a pair of 3D disks settling in a viscous fluid are much less understood. In this work, a Lattice Boltzmann Method (LBM) and a cylindrical particle Discrete Element Method (DEM) are combined to simulate sedimentations of a pair of disks in a Newtonian viscous fluid for a comprehensive understanding of such processes. Corresponding physical experiments are also conducted to verify the simulation results and to complement the investigations. The disks of various dimensionless moments of inertia $I^*$ are used in the simulations and experiments. The falling motions of the disks are analyzed and classified for wide ranges of $I^*$ (0.006 - 0.14) and the Reynolds numbers $Re$ (5 - 1000). Phase diagrams are eventually drawn to determine falling patterns and contact styles of the two interactive disks based on $I^*$ and $Re$.

## 2. Methodology

In the coupled LBM-DEM method, the flow field is solved by the LBM, which was developed by Wang et al. (2016) based on the lattice Boltzmann method coupled with a proper treatment of the moving fluid-solid interface (Peng et al., 2016). The dynamics and interactions of the disks are modeled by the cylindrical particle DEM, which was implemented by Guo et al. (2012b). The momentum exchange method (Wen et al., 2014) is used to calculate the fluid-disk interaction forces with the disk surfaces treated as no-slip boundaries by an interpolated bounce-back scheme (Bouzidi et al., 2001).

### 2.1 Lattice Boltzmann Method

The multi-relaxation time (MRT) LBM (D'Humieres et al., 2002) is used to solve the evolution of mesoscopic velocity distribution function,

$$\mathbf{f}(\mathbf{x} + \mathbf{e}_\alpha \delta_t, t + \delta_t) = \mathbf{f}(\mathbf{x}, t) - \mathbf{M}^{-1} \cdot \mathbf{S} \cdot [\mathbf{m} - \mathbf{m}^{(eq)}] + \mathbf{Q}\delta_t, \tag{2.1}$$

where **f** is the particle distribution function for the discrete velocity $\mathbf{e}_\alpha$ in the $\alpha$th direction at the position **x** and time $t$, $\delta_t$ is the lattice time step, **M** is an orthogonal transformation matrix that relates the distribution function **f** and moment space **m** as $\mathbf{m} = \mathbf{M} \cdot \mathbf{f}$ and $\mathbf{f} = \mathbf{M}^{-1} \cdot \mathbf{m}$, $\mathbf{m}^{(eq)}$ represents the equilibrium distribution of the moment space **m**, and **S** is the diagonal relaxation matrix which specifies relaxation rates for nonconservative moments. Forcing field **Q** (Guo et al., 2002) applies a body force to the fluid and the $\alpha$th component of **Q** is calculated as,

$$Q_\alpha = \omega_\alpha \left[ \frac{\mathbf{e}_\alpha \cdot \mathbf{F}}{c_s^2} + \frac{\mathbf{Fu}:(\mathbf{e}_\alpha \mathbf{e}_\alpha - c_s^2 \mathbf{I})}{c_s^4} \right], \tag{2.2}$$

in which $\omega_\alpha$ represents a weight coefficient along the $\alpha$th direction, **F** is an external force vector. **u** is the macroscopic fluid velocity vector, the sound speed $c_s$ is equal to $1/\sqrt{3}$ in lattice unit, and **I** is an identity matrix.

Fluid density $\rho$, pressure $p$, and velocity vector **u** are the macroscopic variables, which can be obtained from the mesoscopic distribution function **f**. For the incompressible flows, they have the forms (He and Luo, 1997),

$$\rho = \rho_0 + \delta\rho, \tag{2.3}$$

in which

$$\rho_0 = 1 \text{ and } \delta\rho = \sum_\alpha f_\alpha,$$

$$p = \delta\rho c_s^2, \tag{2.4}$$

$$\rho_0 \mathbf{u} = \sum_\alpha f_\alpha \mathbf{e}_\alpha + \frac{\delta_t}{2} \rho_0 \mathbf{F}. \tag{2.5}$$

As showed in Figure 1, nineteen-velocity model in three dimensions, i.e., D3Q19 model, is adopted for the discrete velocities, which are written as,

$$\mathbf{e}_\alpha = \begin{cases} (0,0,0), & \alpha = 0, \\ (\pm 1,0,0), (0,\pm 1,0), (0,0,\pm 1), & \alpha = 1,2,\dots,6, \\ (\pm 1,\pm 1,0), (\pm 1,0,\pm 1), (0,\pm 1,\pm 1), & \alpha = 7,8,\dots,18. \end{cases} \tag{2.6}$$

Thus, the weight coefficient $\omega_\alpha$ is determined as,

$$\omega_\alpha = \begin{cases} 1/3, & e_\alpha^2 = 0 \\ 1/18, & e_\alpha^2 = 1 \\ 1/36, & e_\alpha^2 = 2 \end{cases}. \tag{2.7}$$

The expressions of **M**, **m**, **m**$^{(eq)}$, and **S** of the D3Q19 model are provided in the work by Wang et al. (2016).

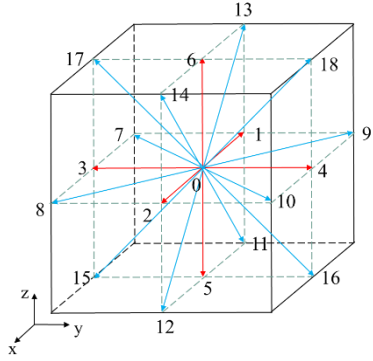

Figure 1: An illustration of D3Q19 model.

### 2.1.1 No-slip boundary condition

The no-slip condition on solid surfaces is crucial for interface-resolved particle-laden flow simulations. In the present LBM, an interpolated bounce-back scheme of second-order accuracy, proposed by Bouzidi et al. (2001), is used to account for the no-slip boundaries. As shown in Figure 2, $\mathbf{x}_w$ is the location where a boundary link intercepts with the solid surface, $\mathbf{x}_f$ and $\mathbf{x}_b$ are the locations of two lattice nodes on the opposite sides of the solid boundary. The solid boundary location is described by $q = |\mathbf{x}_f - \mathbf{x}_w|/|\mathbf{x}_f - \mathbf{x}_b|$. When $q \leq 0.5$, the distribution function at $\mathbf{x}_r$ (located between $\mathbf{x}_f$ and $\mathbf{x}_{ff}$) is first interpolated and then streamed exactly to $\mathbf{x}_f$ after bouncing back from the wall for a total distance of a lattice grid size $\delta_x$ in a single time step $\delta_t$, as shown in Figure 2a. Therefore, the velocity distribution function at the boundary node is updated using a linear-interpolation scheme,

$$f_\alpha(\mathbf{x}_f, t + \delta_t) = 2q f^*_{\bar{\alpha}}(\mathbf{x}_f, t) + (1 - 2q) f^*_{\bar{\alpha}}(\mathbf{x}_{ff}, t) + 2\rho_0 \omega_\alpha \frac{\mathbf{e}_\alpha \cdot \mathbf{u}_w}{c_s^2}, \quad (2.8)$$

or a quadratic-interpolation one,

$$f_\alpha(\mathbf{x}_f, t + \delta_t) = q(2q + 1) f^*_{\bar{\alpha}}(\mathbf{x}_f, t) + (1 + 2q)(1 - 2q) f^*_{\bar{\alpha}}(\mathbf{x}_{ff}, t) -$$
$$q(1 - 2q) f^*_{\bar{\alpha}}(\mathbf{x}_{fff}, t) + 2\rho_0 \omega_\alpha \frac{\mathbf{e}_\alpha \cdot \mathbf{u}_w}{c_s^2}, \quad (2.9)$$

where $f_\alpha$ and $f^*_{\bar{\alpha}}$ are the bounce-back distribution function and incident distribution function, respectively, with $\mathbf{e}_\alpha = -\mathbf{e}_{\bar{\alpha}}$, and $\mathbf{u}_w$ is the velocity at the wall location $\mathbf{x}_w$. When $q > 0.5$, as shown in Figure 2b, the distribution function at $\mathbf{x}_f$ is first streamed to $\mathbf{x}_r$ (located between $\mathbf{x}_f$ and $\mathbf{x}_w$) after bouncing back from the wall, then an interpolation is performed to obtain $f_\alpha(\mathbf{x}_f, t + \delta_t)$, yielding a linear scheme,

$$f_\alpha(\mathbf{x}_f, t + \delta_t) = \frac{1}{2q}\left[f_{\bar{\alpha}}^*(\mathbf{x}_f, t) + 2\rho_0 \omega_\alpha \frac{\mathbf{e}_\alpha \cdot \mathbf{u}_w}{c_s^2}\right] + \frac{(2q-1)}{2q} f_\alpha(\mathbf{x}_{ff}, t + \delta_t), \qquad (2.10)$$

or a quadratic one,

$$f_\alpha(\mathbf{x}_f, t + \delta_t) = \frac{1}{q(2q+1)}\left[f_{\bar{\alpha}}^*(\mathbf{x}_f, t) + 2\rho_0 \omega_\alpha \frac{\mathbf{e}_\alpha \cdot \mathbf{u}_w}{c_s^2}\right] + \frac{2q-1}{q} f_\alpha(\mathbf{x}_{ff}, t + \delta_t) -$$
$$\frac{2q-1}{1+2q} f_\alpha(\mathbf{x}_{fff}, t + \delta_t). \qquad (2.11)$$

If the three fluid nodes $\mathbf{x}_f$, $\mathbf{x}_{ff}$, and $\mathbf{x}_{fff}$ are all available, the quadratic-interpolation schemes (Eqs. 2.9 and 2.11) are used. If only two fluid nodes $\mathbf{x}_f$ and $\mathbf{x}_{ff}$ exist ($\mathbf{x}_{fff}$ is absent or separated from $\mathbf{x}_{ff}$ by a solid boundary), the linear schemes (Eqs. 2.8 and 2.10) should be invoked.

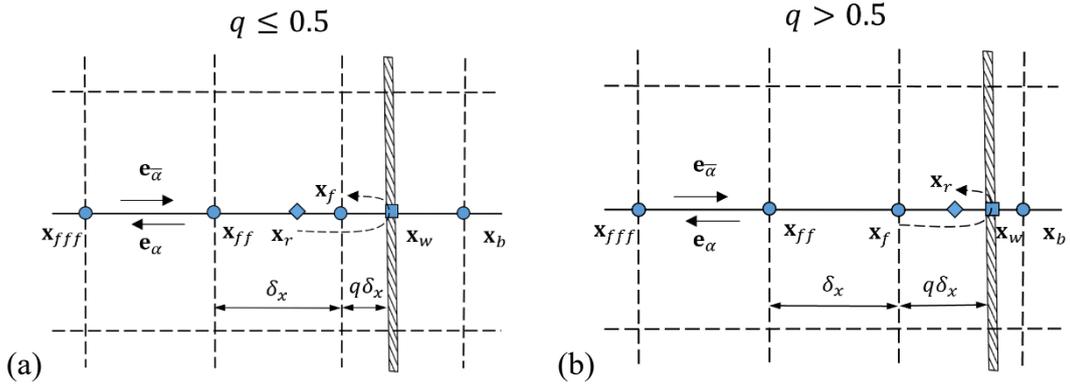

Figure 2: The bounce-back scheme for the no-slip boundary condition near a solid wall.

### 2.1.2 Force evaluation

Based on the work by Ladd (1994), Wen et al. (2014) proposed a Galilean invariant momentum exchange method (GIMEM) to calculate hydrodynamic force $\mathbf{F}_H$ and torque $\mathbf{T}_H$ exerted on a solid object in a fluid,

$$\mathbf{F}_H \delta_t = \sum_{\mathbf{x}_f, \alpha}\left[f_{\bar{\alpha}}^*(\mathbf{x}_f, t)(\mathbf{e}_{\bar{\alpha}} - \mathbf{u}_w) - f_\alpha(\mathbf{x}_f, t + \delta_t)(\mathbf{e}_\alpha - \mathbf{u}_w)\right], \qquad (2.12)$$

$$\mathbf{T}_H \delta_t = \sum_{\mathbf{x}_f, \alpha}(\mathbf{x}_w - \mathbf{x}_c) \times \left[f_{\bar{\alpha}}^*(\mathbf{x}_f, t)(\mathbf{e}_{\bar{\alpha}} - \mathbf{u}_w) - f_\alpha(\mathbf{x}_f, t + \delta_t)(\mathbf{e}_\alpha - \mathbf{u}_w)\right], \qquad (2.13)$$

in which $\mathbf{x}_c$ is the position of mass center of the solid object. The summation is performed over all boundary links $\alpha$ and all boundary nodes $\mathbf{x}_f$.

In the present LBM simulations, the solid disks move within the fixed lattice grids. Thus, when a grid node is uncovered as the solid moves, proper initial distribution functions should be assigned to this node to ensure that the flow is properly defined at the new fluid node. An 'equilibrium plus non-equilibrium' refilling scheme originally proposed by Caiazzo (2008) is used in the present work. In this refilling scheme, the distribution functions at a newly uncovered node $\mathbf{x}_{uc}$ are partitioned into the equilibrium and non-equilibrium parts,

$$f_\alpha(\mathbf{x}_{uc}, t + \delta_t) = f_\alpha^{(eq)}(\mathbf{u}_w, \overline{\delta\rho}) + f_\alpha^{(neq)}(\mathbf{x}_{uc} + \mathbf{e}_c\delta_t, t + \delta_t) . \tag{2.14}$$

The equilibrium part $f_\alpha^{(eq)}$ is obtained from the local wall velocity $\mathbf{u}_w$ and the average density fluctuation of available neighboring fluid nodes $\overline{\delta\rho}$. The non-equilibrium term $f_\alpha^{(neq)}$ is determined as the non-equilibrium part from the neighboring node $\mathbf{x}_{uc} + \mathbf{e}_c\delta_t$, in which $\mathbf{e}_c$ is the lattice direction giving the minimum value of $(\mathbf{e}_c \cdot \mathbf{n})/(|\mathbf{e}_c||\mathbf{n}|)$ and $\mathbf{n}$ is the outer normal vector on the solid surface from where the new fluid node is uncovered. According to the comparative studies among different refilling schemes (Peng et al., 2016), the present 'equilibrium plus non-equilibrium' refilling scheme performed well in computational efficiency and numerical stability.

As two solid cylindrical objects come close to each other with a small gap between them, the lattice grids are not sufficiently fine to resolve the fluid dynamics within the gap. Thus, the lubrication force model proposed by Brändle de Motta et al. (2013) is used to calculate the additional hydrodynamic forces exerted on the cylinders due to the squeezed thin fluid layer,

$$F_L^{ij}(\varepsilon) = -6\pi\mu_f R^* v_n [\lambda(\varepsilon) - \lambda(\varepsilon_0)] , \tag{2.15}$$

in which $\mu_f$ is the fluid viscosity and $v_n$ is the normal component of the relative velocity between the two cylinders at the approaching point. The effective solid size $R^* = R_c^i R_c^j / (R_c^i + R_c^j)$, in which $R_c^i$ and $R_c^j$ are the characteristic sizes of the two cylinders defined as the minimum value between the radius $r_c$ and the half length $l_c/2$ of the cylinder, i.e. $R_c = \text{MIN}(r_c, l_c/2)$. For the collision between a cylinder and a planar wall, the effective solid size is equal to the characteristic size of the cylinder $R^* = R_c^i$. The asymptotic functions $\lambda(\varepsilon)$, in which $\varepsilon = \delta_{gap}/R_c$ and $\delta_{gap}$ is the gap size between two approaching cylinders, have different forms for the cylinder-cylinder and cylinder-wall interactions (Brändle de Motta et al., 2013),

$$\text{cylinder-cylinder:} \quad \lambda(\varepsilon) = \frac{1}{2\varepsilon} - \frac{9}{20}\ln\varepsilon - \frac{3}{56}\varepsilon\ln\varepsilon + 1.346 , \tag{2.16a}$$

$$\text{cylinder-wall:} \quad \lambda(\varepsilon) = \frac{1}{\varepsilon} - \frac{1}{5}\ln\varepsilon - \frac{1}{21}\varepsilon\ln\varepsilon + 0.9713. \tag{2.16b}$$

As shown in Figure 3, three critical scaled gap sizes $\varepsilon_0$, $\varepsilon_1$, and $\varepsilon_2$ are defined. The lubrication force is invoked when $\varepsilon < \varepsilon_0$. The lubrication force $F_L^{ij}(\varepsilon)$ is calculated using Eqs. (2.15) and (2.16a or b) when $\varepsilon_1 < \varepsilon < \varepsilon_0$, and it is constant and equal to the value at $\varepsilon = \varepsilon_1$, i.e. $F_L^{ij}(\varepsilon) = F_L^{ij}(\varepsilon_1)$, when $0 \leq \varepsilon \leq \varepsilon_1$. Solid-solid contact between the two cylinders occurs when $\varepsilon < 0$, then the solid-solid contact force $F_{ss}^{ij}$ should be considered,

$$F_{ss}^{ij} = -k_n|\delta_{gap}| - \beta_n v_n, \tag{2.17}$$

$$k_n = -\frac{m_e(\pi^2 + (\ln e_d)^2)}{(N_c \delta_t)^2}, \tag{2.18}$$

$$\beta_n = -\frac{2m_e(\ln e_d)}{(N_c \delta_t)}, \tag{2.19}$$

where $m_e = m_i m_j/(m_i + m_j)$ for the cylinder-cylinder collision and $m_e = m_i$ for the cylinder-wall collision with the masses of the cylinders $m_i$ and $m_j$. The coefficient of restitution is specified as $e_d = 0.97$, and $N_c \delta_t$ represents the contact duration in which $N_c$ is set to 8 in the present simulations as suggested by Brändle de Motta et al. (2013). The interaction force is calculated as a sum of the solid-solid contact force $F_{ss}^{ij}$ and lubrication force $F_L^{ij}(\varepsilon_1)$ when $\varepsilon_2 \leq \varepsilon < 0$, and only the solid-solid contact force exists when $\varepsilon \leq \varepsilon_2$. In the present simulations, the three critical gap sizes $\varepsilon_0$, $\varepsilon_1$, and $\varepsilon_2$ are specified as 0.125, 0.001, and -0.005, respectively, for the cylinder-cylinder contact, and 0.15, 0.001, and -0.005, respectively, for the cylinder-wall contact. It is noted that the lubrication force models represented by Eqs.(2.15) and (2.16a and b), which were originally calibrated for the sphere-sphere and sphere-wall interactions, are used for the interactions involving the cylinders in the present work. The effect of this approximate treatment is considered small, because the magnitudes of the lubrication forces are much smaller than those of resolved hydrodynamic forces, as illustrated below by the results in Sec. 5.

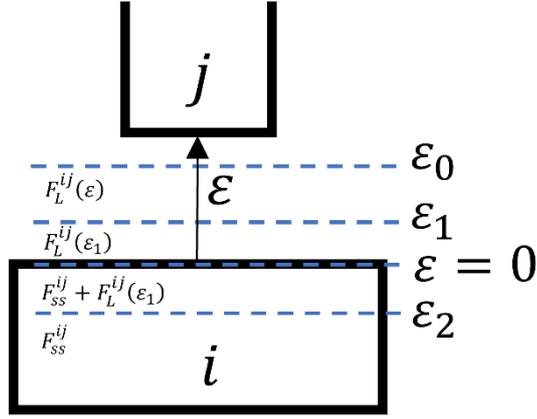

Figure 3: Multilayer lubrication force model for two contacting cylindrical objects.

## 2.2 Discrete element method for cylindrical particles

The code of the cylindrical particle Discrete Element Method (DEM) used in the present study was originally developed by Guo et al. (2012a; 2012b). The translational and rotational motion of an individual cylinder *i* is governed by Newton's second law of motion,

$$m_i \frac{\mathrm{d}\mathbf{v}_i}{\mathrm{d}t} = \mathbf{F}_H^i + \mathbf{F}_L^i + \mathbf{F}_{ss}^i + m_i \mathbf{g}, \qquad (2.20)$$

and

$$\mathbf{I}_i \cdot \frac{\mathrm{d}\boldsymbol{\omega}_i}{\mathrm{d}t} - (\mathbf{I}_i \cdot \boldsymbol{\omega}_i) \times \boldsymbol{\omega}_i = \mathbf{T}_H^i + \mathbf{T}_L^i + \mathbf{T}_{ss}^i, \qquad (2.21)$$

in which, $\mathbf{v}_i$ and $\boldsymbol{\omega}_i$ are the translational and angular velocities, respectively, of the cylinder, and their derivatives with respect to the time *t* give the corresponding accelerations. The translational movement of the cylinder of mass $m_i$ is driven by a combination of the hydrodynamic force $\mathbf{F}_H^i$, lubrication force $\mathbf{F}_L^i$, solid-solid contact force $\mathbf{F}_{ss}^i$, and the gravitational force $m_i \mathbf{g}$. The rotation of the cylinder is propelled by a sum of torques $\mathbf{T}_H^i$, $\mathbf{T}_L^i$, and $\mathbf{T}_{ss}^i$, arising from the hydrodynamic force, lubrication force, and solid-solid contact force, respectively. For the three-dimensional non-spherical objects, $\mathbf{I}_i$ represents the moment of inertia tensor. The evolution of velocities, positions and orientations of the objects can be obtained by the time integration of Eqs. (2.20) and (2.21) using a central finite difference scheme with a fixed time step.

In the DEM simulations, several typical cylinder-cylinder contact types were recognized by Kodam et al. (2010), as shown in Figure 4. In the present work, the contact detection algorithms are proposed to determine the gap size $\delta_{gap}$, contact normal direction, and contact point position for each contact scenario, following the previous work by Kodam et al. (2010) and Guo et al. (2012a). Thus, the lubrication force $\mathbf{F}_L^i$, solid-solid contact force $\mathbf{F}_{ss}^i$, and the two torques $\mathbf{T}_L^i$ and $\mathbf{T}_{ss}^i$ due to $\mathbf{F}_L^i$ and $\mathbf{F}_{ss}^i$ can be calculated.

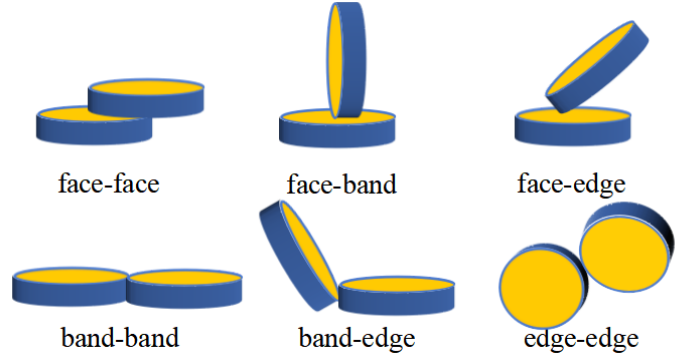

Figure 4: Typical cylinder-cylinder contact types considered in the DEM simulations (Kodam et al., 2010; Guo et al., 2012a).

## 2.3 Validation of the LBM-DEM code

Four sets of simulations are performed and analyzed using the developed LBM-DEM code. Firstly, the sedimentations of a single disk of aspect ratio $AR = 0.1$ and dimensionless moment of inertia $I^* = 5.97 \times 10^{-3}$ at various particle terminal Reynolds numbers $Re$ are simulated. The aspect ratio $AR$ of a disk is defined as the ratio of the thickness $l_c$ to the diameter $d_c$ of the disk, i.e. $AR = l_c/d_c$. In the simulations, steady, transitional, and periodic falling patterns are sequentially obtained as $Re$ increases. The simulations are consistent with the previous experimental observations by Field et al. (1997). Secondly, the drag coefficients of the disk ($AR = 0.1$) falling in a fluid at a range of particle terminal Reynolds numbers $Re$ from 10 to 400 are obtained from the present simulations. The simulation results are in good agreement with the predictions by Clift et al. (1978), which were determined by the extensive experimental results. Thirdly, fluid flows around a cylinder ($AR = 5$) fixed at a specified position are simulated. As the cylinder Reynolds number is specified as $Re = 300$, the lift and torque coefficients of the cylinder at various orientational angles (or attack angles) are compared with the previous simulation results by Zastawny et al. (2012). Also, good agreement is achieved. Fourthly, a simulation of two-disk falling in tandem is performed and compared with the existing experimental results by Brosse & Ern (2011). The details about the LBM-DEM code validation and the definitions of the Reynolds numbers are provided in the Appendix A.

In addition, the contact detection algorithms and contact force models for the cylinder-cylinder and cylinder-wall interactions were validated in our previous work by modeling the cylindrical particle packings in a container (Tangri et al., 2017) and the hopper flows (Tangri et al., 2019).

## 3. Computational setup of the sedimentation of two interacting disks

A three-dimensional rectangular computational domain of dimensions $L_x = L_z = 300$ (lattice unit or LU) and $L_y = 1200$ (LU) is created, as shown in Figure 5a. Two identical disks of the equivalent volume sphere diameter $d_{eq}$ are placed in tandem at the positions of the coordinates $(L_x/2, L_y-d_1, L_z/2)$ and $(L_x/2, L_y - d_1 - d_2, L_z/2)$, respectively. The clearance to the top boundary is specified as $d_1 = 6.27 d_{eq}$. The initial distance between the centers of two disks $d_2$ (Figure 5a) and the initial inclination angle $\theta_0$, which is the angle between the major axis of a disk and the $y$ axis (Figure 5b), are varied to understand the effects of the initial conditions of the disk release. Four different aspect ratios ($AR = 0.1, 0.4, 0.7$ and $1$), defined as the ratio of the thickness $l_c$ to the diameter $d_c$ of the disk, are used in the present simulations to examine the effect of disk shape on the falling dynamics. All the disks have the same volume and the equivalent volume sphere diameter is assigned as $d_{eq} = 31.88$ (LU). The densities of fluid and cylinders are assigned as $\rho_f = 1$ and $\rho_s = 1.2$, respectively. The dimensionless moments of inertia $I^*$, which can be calculated for the disks using Eq.(1.1), are listed in Table 1. The ratios of the moment of inertia of the disk about its axis of symmetry $I_l$ and the moment of inertia of the same disk about its diameter $I_d$ are also shown in Table 1. A ratio of $I_l/I_d$ reflects the relative significance of the inertia rotating about the axis of symmetry to that about a diameter for a disk.

| AR | $I^*$ | $I_l/I_d$ |
|---|---|---|
| 0.1 | $5.97 \times 10^{-3}$ | 1.97 |
| 0.4 | $2.86 \times 10^{-2}$ | 1.65 |
| 0.7 | $6.88 \times 10^{-2}$ | 1.20 |
| 1 | $1.37 \times 10^{-1}$ | 0.857 |

Table 1: Moments of inertia of the disks used in the simulations.

No-slip wall boundary conditions are specified in the $x$ and $z$ directions. Periodic boundary conditions are used in the $y$ direction for the modeling of the disks falling in a very deep tank. The fluid and two disks are initially at rest, and sedimentation starts when constant downward body forces are exerted on the disks. The disks settle at different terminal velocities by adjusting the fluid viscosity. In the present simulations, the confinement effects of solid wall boundaries are minimized by using a sufficiently large domain. The size ratio of $L_x/d_{eq}$ and $L_z/d_{eq}$ is set to about 9.4, resulting in nearly the same results of settling paths and rotation of the disks as using a wider domain of the ratio $L_x/d_{eq} = L_z/d_{eq} = 18.8$. The effect of height $L_y$, the distance between two periodic boundaries, is examined by increasing the ratio $L_y/d_{eq}$ from 37.6 to 75.3 for the simulations with $I^* = 5.97 \times 10^{-3}$ and $Re = 600$, and nearly the same results are obtained. Thus, a smaller domain of $L_x/d_{eq} = L_z/d_{eq} = 9.4$ and $L_y/d_{eq} = 37.6$ is used in most of the present simulations for a lower computational cost. The sensitivity to the grid resolution has been examined by doubling the present grid resolution, and almost the identical simulation results are obtained.

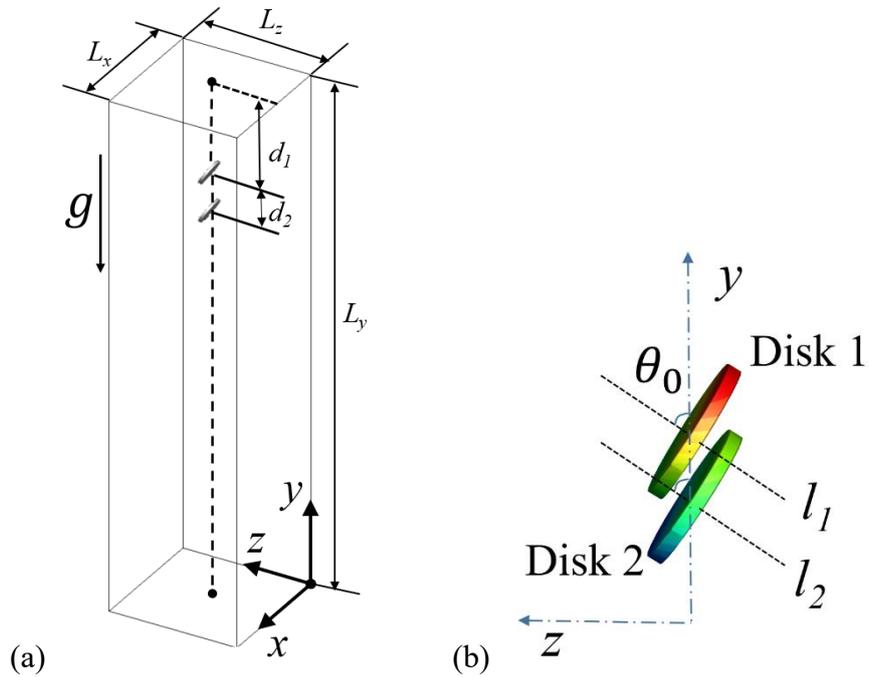

Figure 5: (a) Computational domain of two disks falling in a viscous fluid and (b) an illustration of the initial angle between the major axis of a disk and the vertical direction $y$.

## 4. Experimental setup of the sedimentation of two interactive disks

Physical experiments of two interacting disks falling in a fluid are conducted to provide additional results for a better understanding of sedimentation behaviors and to validate the present simulation results. Six types of disks with different sizes (diameter $d_c$ and thickness $l_c$) and aspect ratios ($AR$), as listed in Table 2, are used in the experiments. The disks, made of acrylic of density $\rho_s = 1.2$ g/cm$^3$, are manufactured using a laser cutting machine. To achieve different settling Reynolds numbers, various solutions of glycerol in a water with different viscosities are employed, as shown in Table 3.

The experiments are conducted at the room temperature of 20°C. At the beginning, two identical disks are held by a single clamp and are fully immersed in the stationary liquid inside a tank of 200 × 2000 × 200 mm$^3$. The initial positions and orientations of the two disks are the same to those specified in the present numerical simulations. The clamp opens slowly and slightly to release the two disks simultaneously, minimizing flow perturbations. The two disks then settle in the liquid under the effect of gravitational force. The falling processes of the disks are recorded by cameras, and the videos are analyzed by a MATLAB® code to determine the positions and settling speeds of the disks. In the present experiments, at least three runs are conducted for a specified set of control parameters.

| No. | $AR$ | $d_c$ /mm | $l_c$ /mm |
| --- | --- | --- | --- |
| 1 | 0.1±0.0015 | 20±0.12 | 2±0.02 |
| 2 | 0.1±0.0011 | 28±0.15 | 2.8±0.025 |
| 3 | 0.4±0.0022 | 7±0.02 | 2.8±0.02 |
| 4 | 0.7±0.0032 | 10±0.08 | 7±0.06 |
| 5 | 1±0.013 | 2.8±0.02 | 2.8±0.01 |
| 6 | 1±0.008 | 10±0.09 | 10±0.07 |

Table 2: Geometric properties of six types of disks.

| Mass concentration of glycerol | 0 | 25% | 50% |
| --- | --- | --- | --- |
| Density of mixture (g/cm³) | 1 | 1.0598 | 1.1263 |
| Dynamic viscosity of mixture (mPa·s) | 1.005 | 2.095 | 6.05 |
| Kinematic viscosity of mixture (μm/s²) | 1.005 | 1.9768 | 5.3714 |

Table 3: Properties of three solutions at the temperature of 20°C.

## 5. Results and discussion

To understand the sedimentation process, we analyze horizontal drifting displacements, orientation, and hydrodynamic forces of the two disks. The orientation of a disk can be described by a pitch angle $\theta$ and a yaw $\phi$. A shown in Figure 6a, the pitch angle $\theta$ is the angle from the vertical $y$ axis to the line $l'$, which is the projection of the major axis of the disk ($l$) onto the $y$-$z$ plane, and the yaw angle $\phi$ is the angle from $l'$ to $l$.

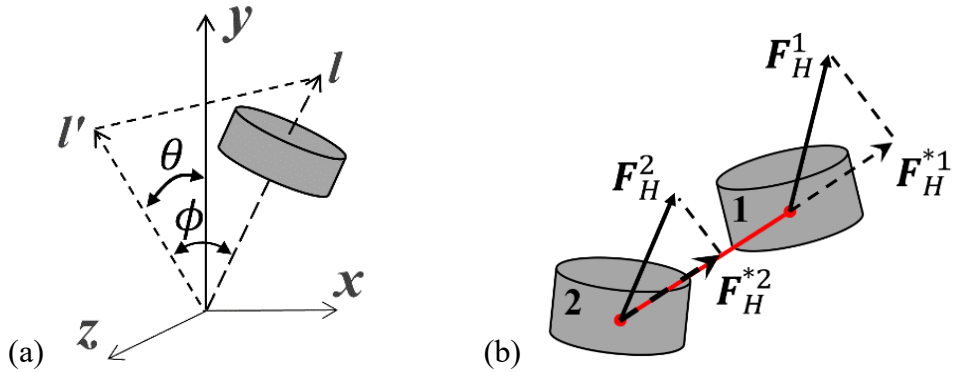

Figure 6: Illustration of (a) pitch angle $\theta$ and yaw angle $\phi$ of a disk and (b) hydrodynamic force components in the direction of the vector connecting two centers of the disks.

The hydrodynamic forces exerted on the disks by surrounding fluid play a critical role in the contacts between the disks. For a better understanding of such hydrodynamic contributions to the disk-disk contacts, the components of the hydrodynamic forces in the direction of the vector connecting the mass centers of the two disks are examined. As illustrated in Figure 6b, the hydrodynamic force component $F_H^{*i}$ can be written as

$$F_H^{*i} = \frac{\boldsymbol{F}_H^i \cdot \boldsymbol{n}_{ji}}{(\rho_s - \rho_f)gV_d} \qquad (i, j = 1, 2), \qquad (5.1)$$

in which 1 and 2 represent the trailing disk and leading disk, respectively, $\boldsymbol{F}_H^i$ is the hydrodynamic force vector acting on the disk $i$, which is calculated using Eq.(2.12), and $\boldsymbol{n}_{ji}$ is the unit branch vector from the mass center of the disk $j$ to that of the disk $i$. The dot product is normalized by the gravitational force term $(\rho_s - \rho_f)gV_d$, in which $V_d$ is the volume of a disk and $g$ is the gravitational acceleration. Thus, the positive and negative values of $F_H^{*i}$ contribute to separation and approaching, respectively, of the two disks.

In the falling, the disks accelerate from zero velocity to a terminal state, in which the vertical component of the disk velocity $u_y$ fluctuates around a time-average value (see Figure 11). Thus, the Reynolds number is defined as,

$$Re = \frac{u_l d_{eq}}{v_f}, \qquad (5.2)$$

in which $u_l$ is the time-average vertical velocity of the leading disk (in the lower position) at the terminal state, $d_{eq}$ is the equivalent volume sphere diameter, and $v_f$ is the kinematic viscosity of the fluid.

The present disk-surface-resolved DNS simulations and experimental results show that the dynamics of two closely-arranged flat disks in the sedimentation are remarkably different from that of two spheres, attributed to the effect of the object shape. Further analyses reveal that the falling patterns are determined by a combination of disk shape (characterized by a dimensionless moment of inertia $I^*$) and Reynolds number $Re$, as a result of complex multiphase flows involving disk-fluid and disk-disk interactions.

### 5.1 Effects of initial conditions

The effects of the initial disk inclination angle $\theta_0$ and distance between mass centers of the two disks $d_2$ (see Figure 5a) have been examined and the results are shown in Figures 7 and 8. The dimensionless horizontal displacement of a disk is expressed as $Z^* = \frac{z}{d_{eq}}$, in which $z$ is the $z$-displacement of the disk. The evolution of $Z^*$ and the pitch angle $\theta$ is plotted as a function of a dimensionless falling displacement defined as $Y^* = \frac{y_2}{d_{eq}}$, in which $y_2$ is the $y$-displacement of the leading disk.

As shown in Figures 7a and 7b, when the disks fall broadside on at zero initial inclination angle $\theta_0 = 0°$, the instability with the oscillation in the horizontal displacement $Z^*$ occurs when the dimensionless falling displacement $Y^*$ is greater than 50. At small initial inclination angles $\theta_0 = 2°$ and $5°$, the significant oscillating horizontal movement $Z^*$ occurs much earlier after $Y^* = 10$, indicating that a small asymmetry at the beginning of disk falling leads to a quicker development of instability. At larger initial angles $\theta_0 = 30°$, $60°$ and $80°$ (see Figures 7c and 7d), the two disks exhibit significant horizontal oscillations in $Z^*$ immediately after they are released. Significant horizontal departure from the original position ($Z^* = 0$) is observed for the paths with very large initial inclination angles $\theta_0 = 60°$ and $80°$. The periodic oscillating behaviors of the angle $\theta$ are similar to those of the horizontal displacement $Z^*$, as shown in Figures 7e-7h. The angle $\theta$ eventually fluctuates around $\theta = 0°$ regardless of the values of initial inclination angles $\theta_0$.

The experiments by Brosse and Ern (2014) showed that at $Re = 255$, the thin disks of $AR = 0.1$ settled in the oscillatory paths with the inclination angles fluctuating periodically in time. These experimental observations are qualitatively similar to the present simulation results. In addition, the global linear stability analysis by Tchoufag et al. (2014) predicted that a disk settled down in a periodic oscillating path with the set of parameters $\theta_0 = 0°$, $Re = 255$, $AR = 0.1$, and $I^* = 5.97 \times 10^{-3}$, and the present simulation results are consistent with the prediction.

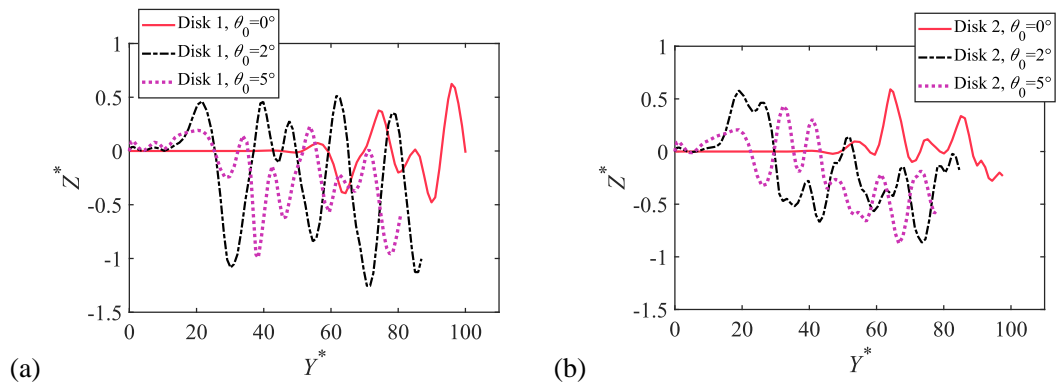

(a)  (b)

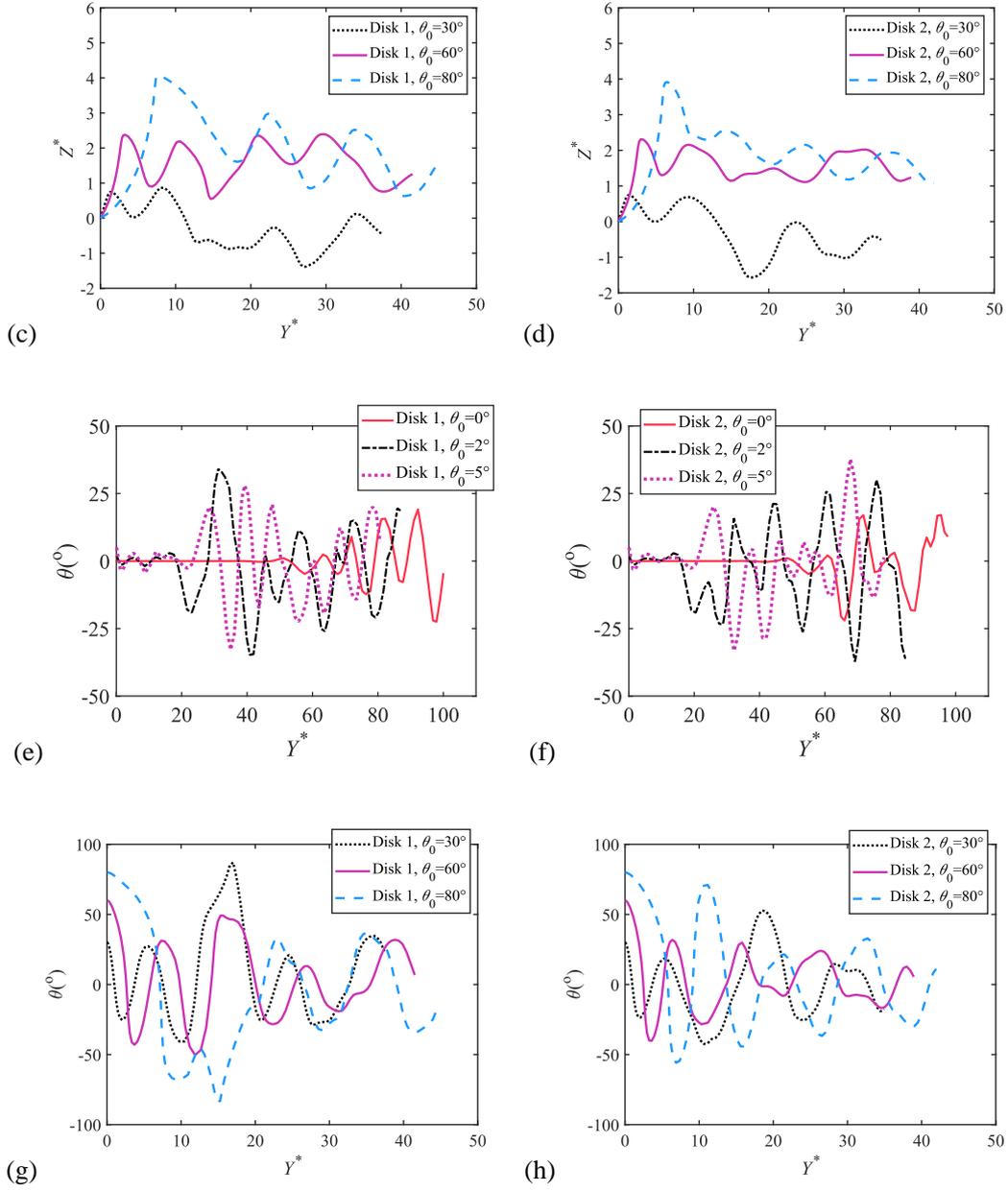

Figure 7: The effects of the initial inclination angle of the disks $\theta_0$ on horizontal displacements in the $z$ direction $Z^*$ (a ~ d) and pitch angles $\theta$ (e ~ h) obtained from the DNS simulations. The results of Disk 1 and Disk 2 are placed in the left-hand and right-hand columns, respectively. The initial distance between the two disks is $d_2 = 2.82 d_{eq}$. The control parameters are $I^* = 5.97 \times 10^{-3}$ and $Re = 255$.

By maintaining the initial inclination angle $\theta_0 = 60°$ and varying the distance $d_2$, the effects of the initial separation between the two disks are shown in Figure 8. For the distances $d_2$ between $0.94 d_{eq}$ and $3.76 d_{eq}$, all the disks oscillate periodically, and the horizontal deviation of the path from the initial position ($Z^* = 0$) increases slightly as the distance $d_2$ increases

(Figures 8a and 8b). Similar periodic oscillations in the pitch angles $\theta$ are observed for all the distances $d_2$ considered (Figures 8c and 8d). Thus, the similar falling patterns are observed for the cases with the initial distance $d_2$ between $0.94d_{eq}$ and $3.76d_{eq}$, though the quantitative differences in the disk motion exist.

In the following sections, in order to study the combined effects of disk aspect ratio and Reynolds number, the initial conditions of the disks are specified as $\theta_0 = 60°$ and $d_2 = 2.82d_{eq}$.

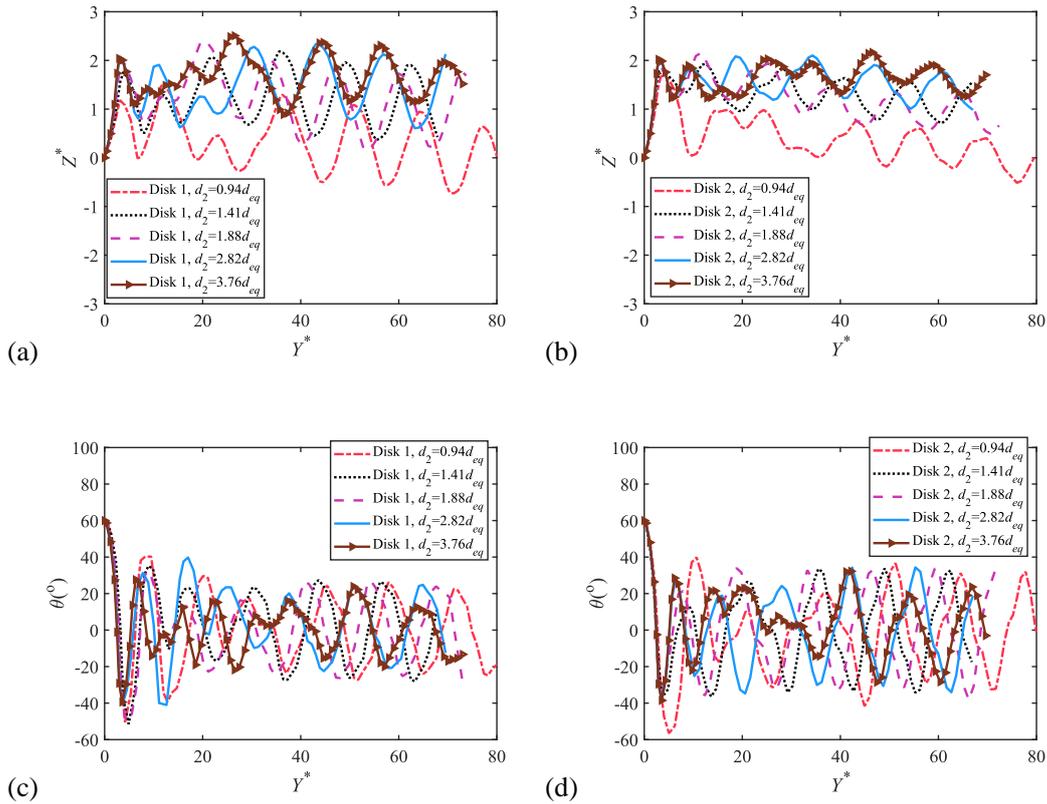

Figure 8: The effects of the initial distance $d_2$ on horizontal displacements in the $z$ direction $Z^*$ (a and b) and pitch angles $\theta$ (c and d) obtained from the simulations. The results of Disk 1 and Disk 2 are placed in the left-hand and right-hand columns, respectively. The initial inclination angle of the disks is $\theta_0 = 60°$. The control parameters are $I^* = 5.97 \times 10^{-3}$ and $Re = 150$.

## 5.2 Steady falling with enduring contact

Sequential snapshots of the two disks of $AR = 0.1$ and $I^* = 5.97 \times 10^{-3}$ falling at a Reynolds number of $Re = 100$ obtained from the simulations are shown in Figure 9. It can be seen that the two falling disks initially translate horizontally in the positive $z$ direction and rotate about the negative $x$ axis, due to the initial inclined angle $\theta_0 = 60°$. After reaching the maximum horizontal displacement in the *z* direction, the disks return to the central vertical line of the domain and meanwhile they rotate about the positive x axis. The trailing disk falls at a faster speed and can collide on the back of the leading disk, since the wake flow behind the leading disk reduces the drag force on the trailing disk. The two disks remain in contact and eventually fall together in a steady mode.

In the steady falling process, the high- and low-pressure regions are formed in the front of and behind, respectively, the two contacting disks (Figure 10a), and the vortices are formed from the edges of the disks (Figure 10b). The high- and low-pressure regions and vortices also remain steady, as the two disks fall in the steady mode. The falling patterns of the two disks obtained from the numerical simulation are similar to those observed from the physical experiment with the same values of $I^*$ and *Re* (Figure 10c). It is noted that the images in Figure 10c are reproduced from the movie recorded in the physical experiment (the movie is provided as a supplementary material). The present steady falling pattern of the two disks in a Y-configuration (Figure 10) was also observed in the previous experiments of two disks falling in tandem with the parameters of $AR = 0.1$ and $Re = 80, 115$ and $152$ (Brosse & Ern, 2011).

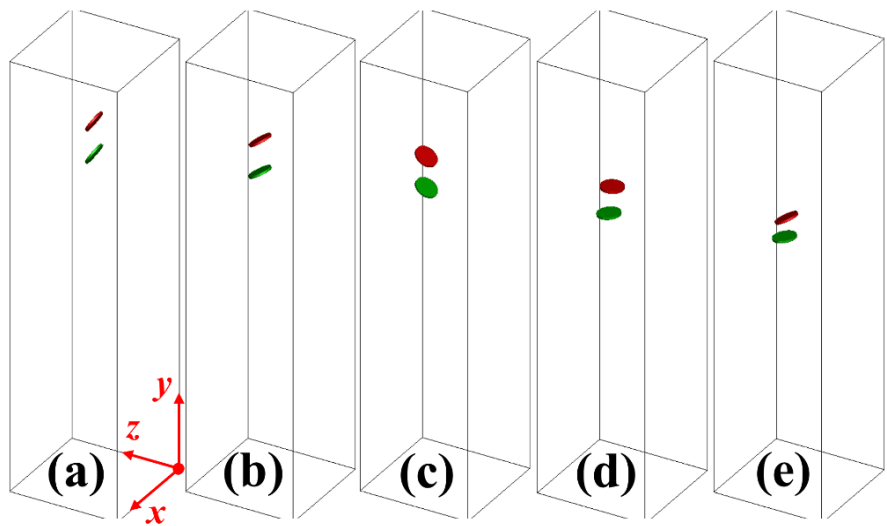

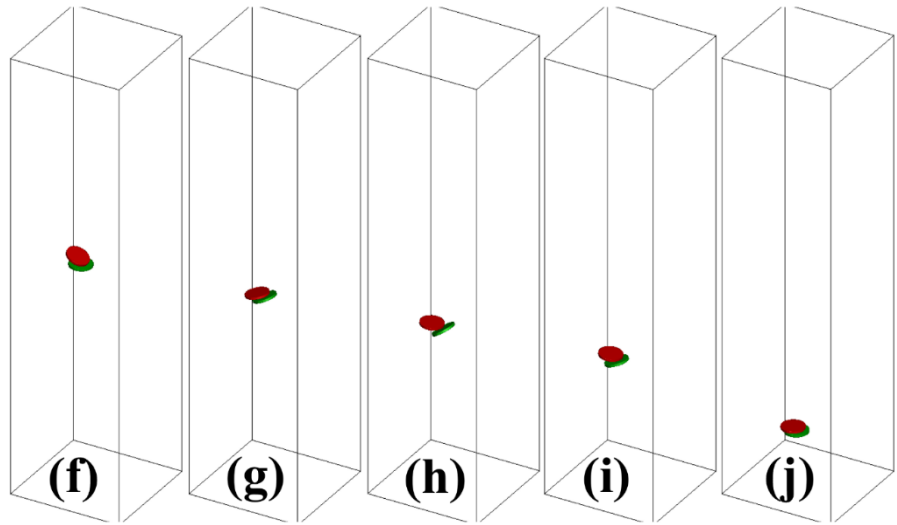

Figure 9: Sequential snapshots of the two disks falling in a fluid obtained from the simulations. The control parameters are $I^* = 5.97 \times 10^{-3}$ and $Re = 100$.

By converting the lattice units to the international system of units (SI), as described in Appendix B, the vertical velocities of the two disks, $u_y$, obtained from the LBM-DEM simulation are compared with those obtained from the real experiment, as shown in Figure 11. In general, the simulation results are in good agreement with the experimental results. At the early stage, the trailing disk 1 has a higher falling speed than the leading disk 2. After the trailing disk catches up with the leading one, the two disks fall at similar speeds. The noisy fluctuations of the experimental data may be attributed to the fact that the resolution of the disk images is not sufficiently high, causing the errors in the determination of the centers of the gravity of the disks.

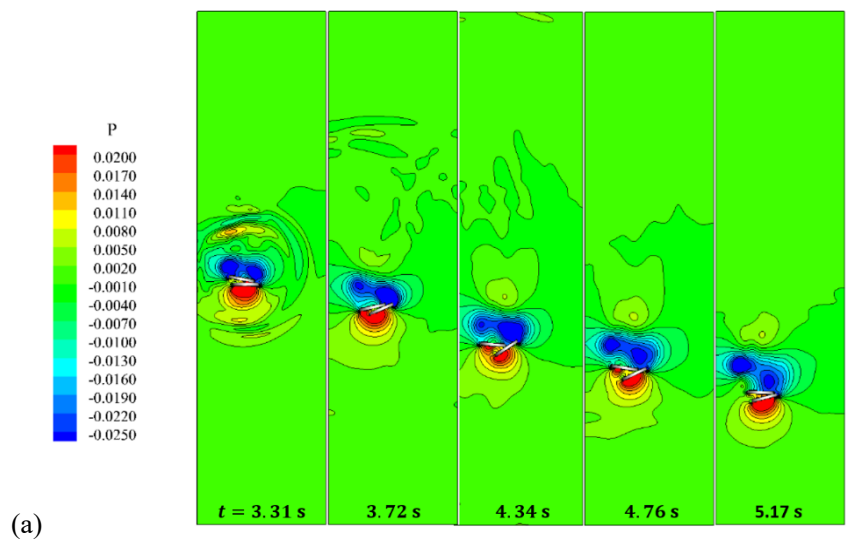

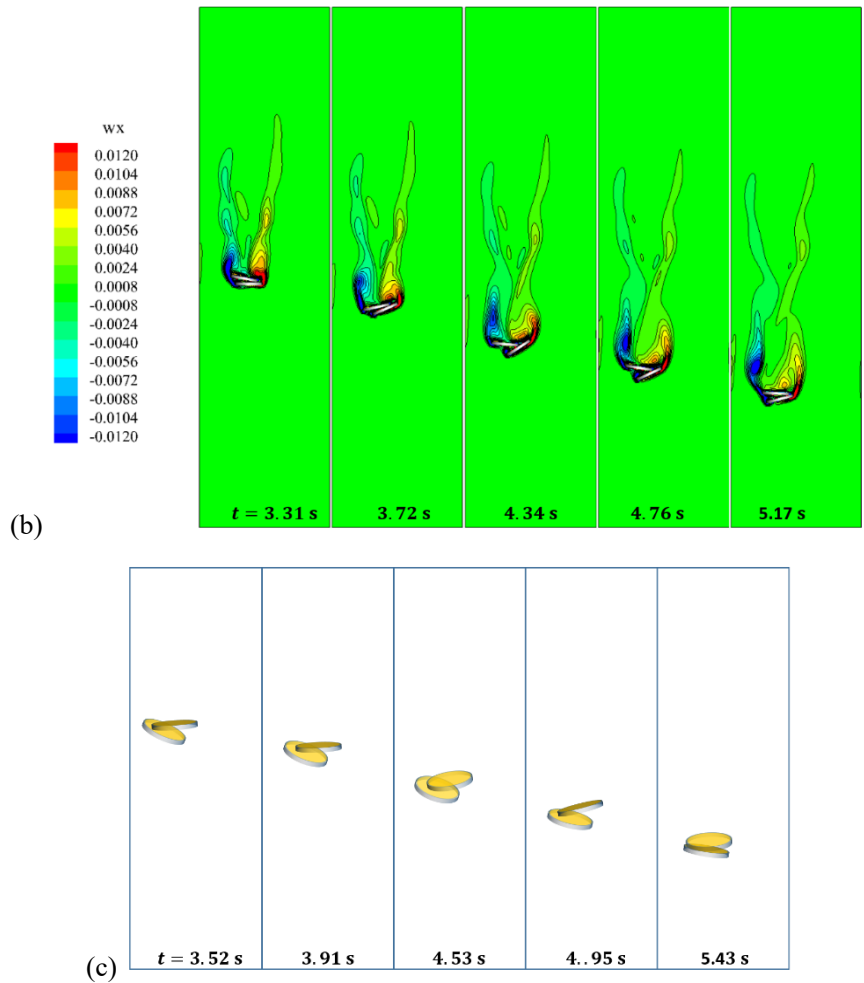

Figure 10: Contours of (a) fluid pressure $P$ and (b) fluid vorticity in the $x$ direction $w_x$ at various time instants $t$ in the sedimentation of two disks obtained from the simulation. The corresponding images at the same time instants obtained from the physical experiment are shown in (c). The views are in the $y$-$z$ plane. The control parameters are $I^* = 5.97 \times 10^{-3}$ and $Re = 100$. The movies of the simulation and physical experiment are provided in Supplementary Materials.

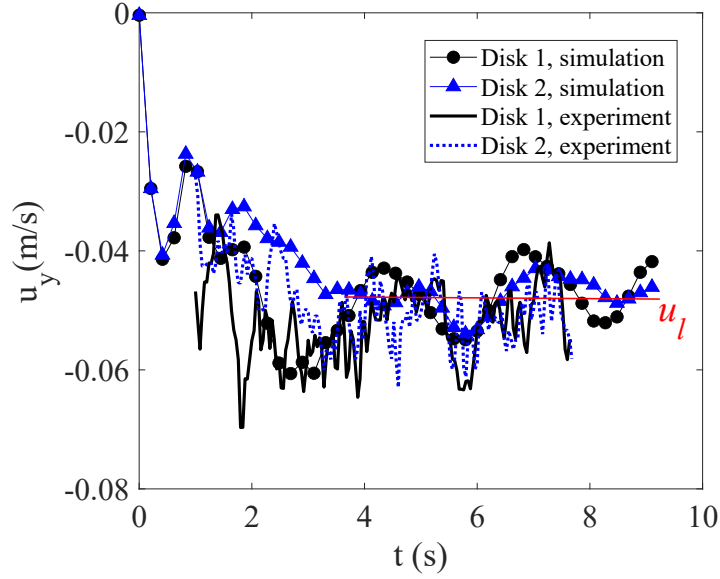

Figure 11: A comparison of the vertical velocities of the disks between the simulation and experiment. The control parameters are $I^* = 5.97 \times 10^{-3}$ and $Re = 100$.

The horizontal displacements of the two disks in the $z$ direction are shown in Figure 12a, in which $Z_1^* = \frac{z_1}{d_{eq}}$, $Z_2^* = \frac{z_2}{d_{eq}}$, $Y^* = \frac{y_2}{d_{eq}}$, $z_1$ and $z_2$ are the $z$-displacements of the trailing disk and leading disk, respectively, and $y_2$ is the $y$-displacement of the leading disk. The two disks initially deviate from the vertical central line ($Z_1^* = Z_2^* = 0$), and then return closer to the central line. The horizontal oscillations are observed in both simulations and experiments. The two disks have initial angles of $\theta_1 = \theta_2 = 60°$ and $\phi_1 = \phi_2 = 0°$, in which the subscripts 1 and 2 represent the trailing and leading disks, respectively. As shown in Figure 12b, the angles $\theta_1$ and $\theta_2$ oscillate periodically with reduced amplitudes. The angles $\phi_1$ and $\phi_2$ maintain 0° at the early stage and then fluctuate with a magnitude smaller than 2° (Figure 12c), indicating that the rotation of the two disks mainly occurs in the $y$-$z$ plane. It is noted that the experimental results in Figures 12a and 12b are from one of three runs under the same conditions. The same falling styles are observed from the three runs, although small differences exist in the displacements of the disks. This also applies to Figures 14a, 14b, and 18a.

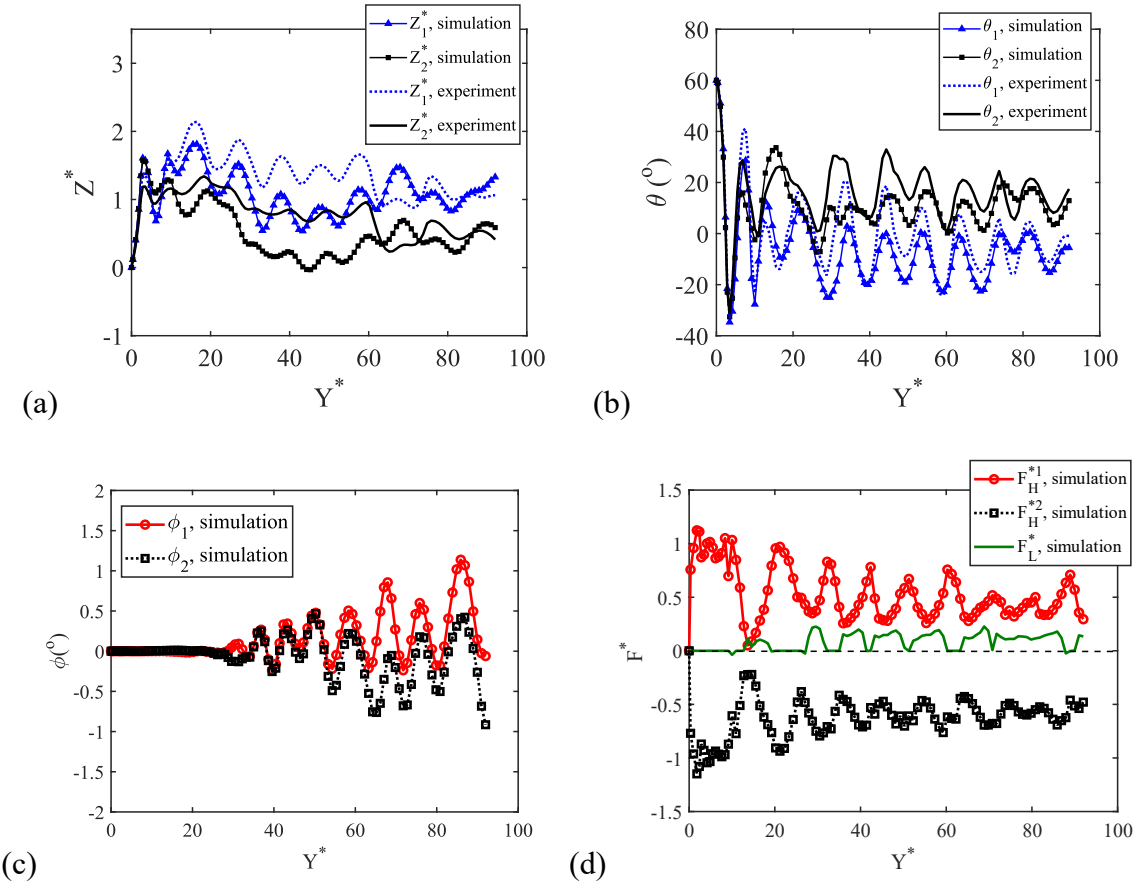

Figure 12: Evolution of (a) horizontal displacements in the $z$ direction, (b) pitch angle $\theta$, (c) yaw angle $\phi$, and (d) hydrodynamic force components and lubrication force. The control parameters are $I^* = 5.97 \times 10^{-3}$ and $Re = 100$.

The evolution of the hydrodynamic force components is plotted in Figure 12d. The hydrodynamic force component on the leading disk $F_H^{*2}$ tends to make the two disks collide, while the hydrodynamic force component on the trailing disk $F_H^{*1}$ separates them. However, the average magnitude of $F_H^{*2}$ is larger than that of $F_H^{*1}$, leading to closing-in and eventually contact of two disks. Consistent with the oscillations of the disk motion (Figures 12a and 12b), the hydrodynamic forces also oscillate periodically. The contact of two disks is defined as the gap between them is sufficiently small (i.e. $\varepsilon < \varepsilon_0$ in Figure 3) that the lubrication force $F_L$ is invoked. The evolution of the scaled lubrication force, $F_L^* = F_L / [(\rho_s - \rho_f) g V_d]$ (the solid-solid contact force $F_{ss}$ is included in $F_L$), is also plotted in Figure 12d, in which a long presence of non-zero values of $F_L^*$ indicates the enduring contact between the two disks.

## 5.3 Periodic swinging with intermittent contact

With the same dimensionless moment of inertia $I^*$ of $5.97 \times 10^{-3}$, the steady falling pattern is changed to double-disk periodic swinging pattern by increasing the Reynolds number $Re$ from 100 to 255. As shown in Figure 13a, the trailing disk can catch up and collide with the leading disk, as the drag force on the trailing disk is reduced by the wake flow behind the leading disk. After the first collision, the trailing disk starts to swing periodically above the leading disk: translational and rotational oscillations occur simultaneously. Similar to the steady falling (Figure 10a), high- and low-pressure regions are generated in the front of and behind, respectively, the two disks, as shown in Figure 13a. Associated with the periodic oscillation of the disks, the vortex shedding occurs periodically, as shown in Figure 13b, which is different from the steady vortices in the steady falling (Figure 10b). The physical experiment of sedimentation is performed with the same parameters of $I^* = 5.97 \times 10^{-3}$ and $Re = 255$ as in the simulation. The snapshots of the disks reproduced from the video taken in the experiment (provided in the supplementary materials) are similar to those obtained from the simulation, as shown in Figure 13c. The periodic oscillations in the two disks falling were also observed in the previous experiments by Brosse & Ern (2011; 2014) with the parameters of $AR = 0.1$ and $Re = 255$ and 275.

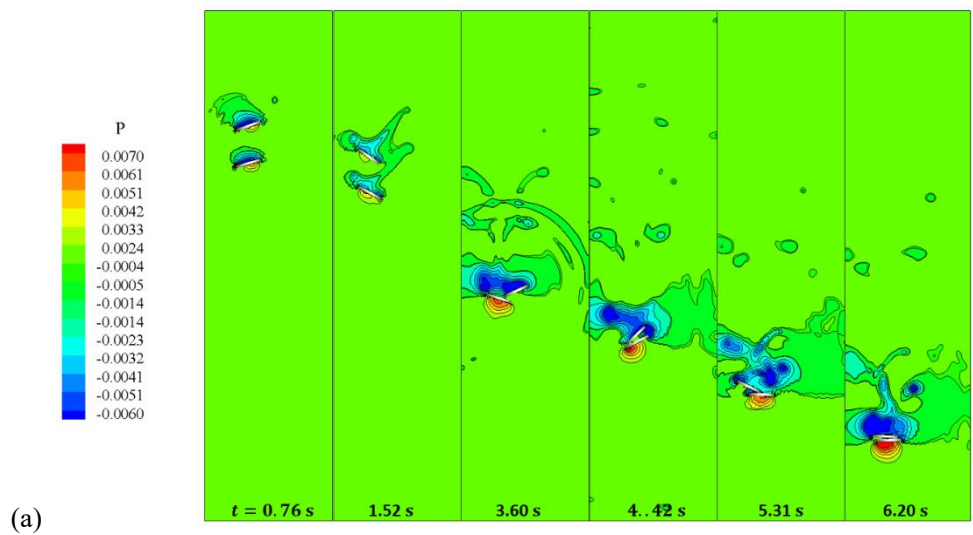
(a)

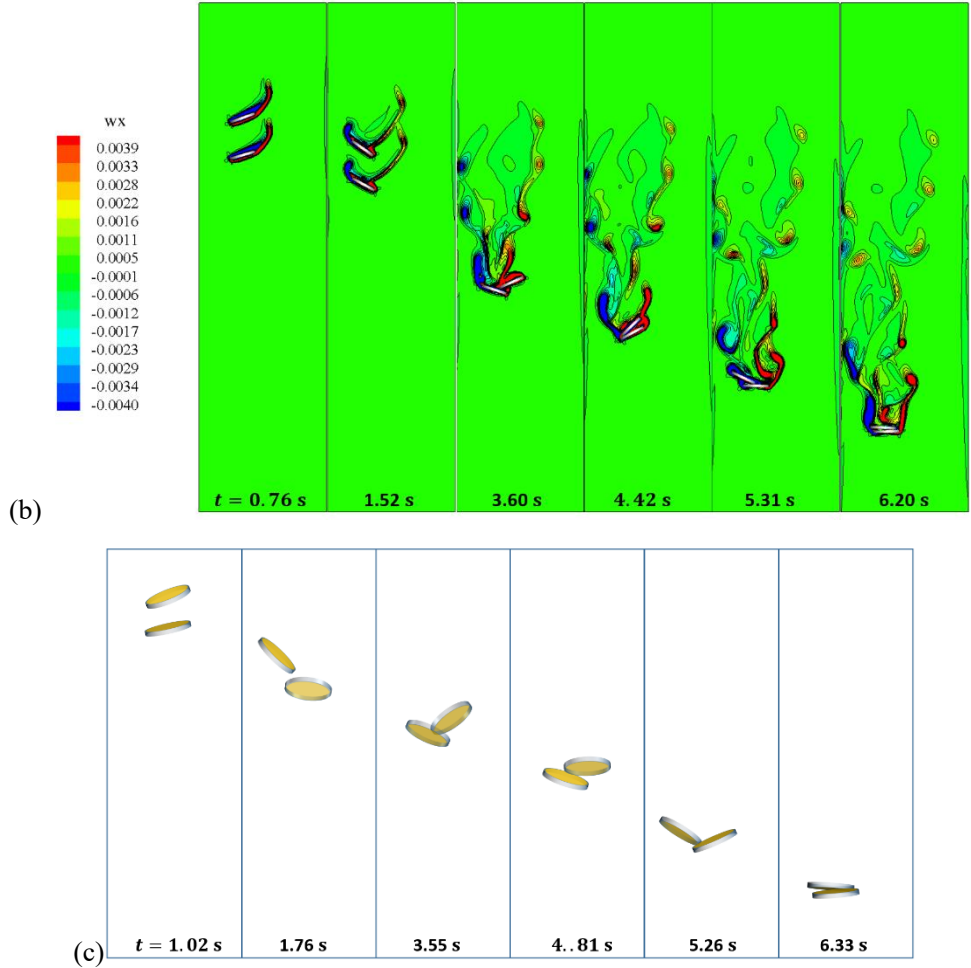

Figure 13: Contours of (a) fluid pressure $P$ and (b) fluid vorticity in the $x$ direction $w_x$ at various time instants in the sedimentation of two disks obtained from the simulation. The corresponding images at the same time instants obtained from the physical experiment are shown in (c). The views are in the $y$-$z$ plane. The control parameters are $I^* = 5.97 \times 10^{-3}$ and $Re = 255$. The movies of the simulation and physical experiment are provided in Supplementary Materials.

In the periodic falling, the translational oscillation in the horizontal $z$ direction and the rotational oscillation are demonstrated in Figures 14a and 14b, respectively. After the early stage, the trailing and leading disks oscillate with a phase difference of about a quarter of period, and the trailing disk has larger oscillating amplitudes than the leading one. Compared to the falling pattern with $Re = 100$ (Figure 12), larger amplitudes of the oscillations in the horizontal displacement $Z^*$ and inclination angle $\theta$ are obtained for the periodic swinging pattern with $Re = 255$. A significant difference between the falling patterns at $Re = 100$ and $Re = 255$ ($I^* = 5.97 \times 10^{-3}$) is that the two disks form a stable Y-configuration at $Re = 100$ (Figures 10b and 10c), and the Y-configuration periodically changes direction at $Re = 255$ (Figures 13b and 13c).

As shown in Figure 14c, the yaw angle $\phi$ develops gradually and remains minimal within $2°$,

indicating the dominance of the planar motion of the disks in the *y-z* plane. When the two disks come to contact, the hydrodynamic force components on them exhibit periodic variations (Figure 14d), which are associated with periodic oscillations of the disks. In the periodic oscillation, the disk-disk contact occurs intermittently as the non-zero lubrication forces $F_L^*$ appear occasionally, as shown in Figure 14d, which is different from the enduring contact in the steady falling (Figure 12d).

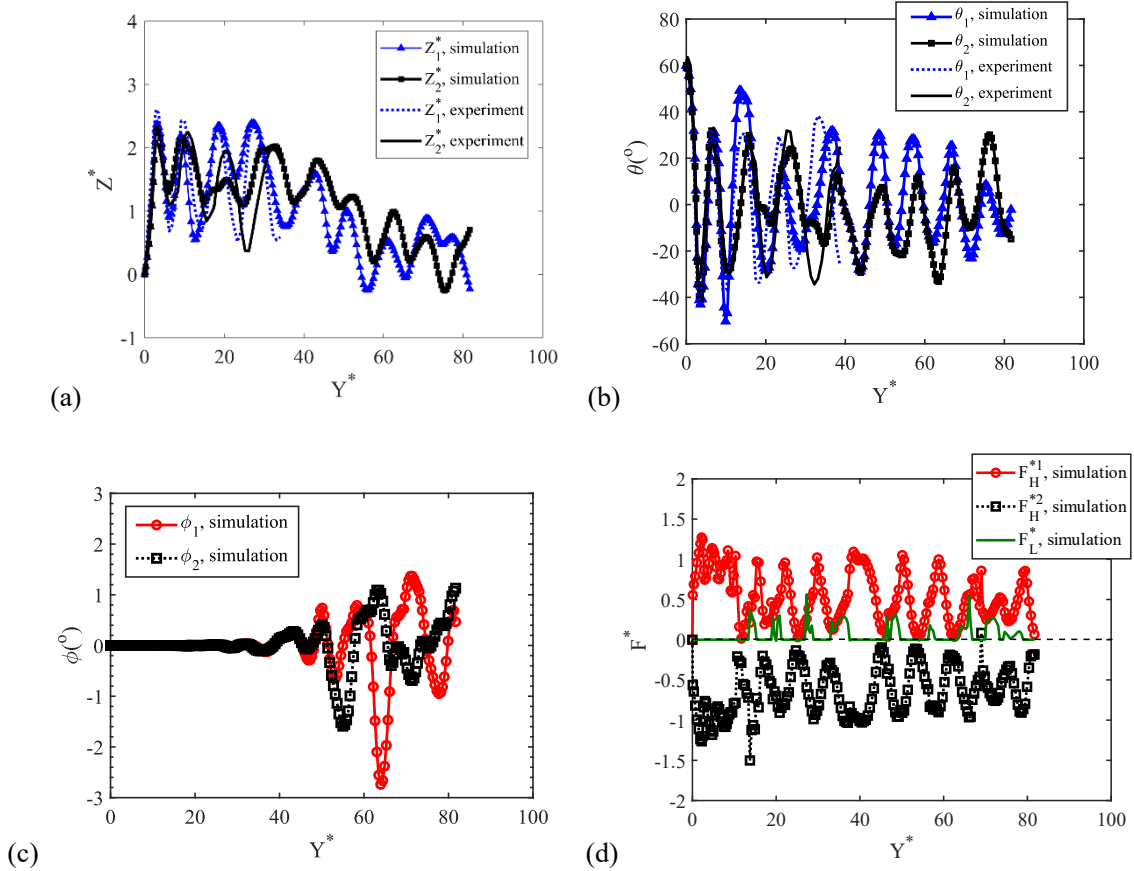

Figure 14: Evolution of (a) horizontal displacements in the *z* direction, (b) pitch angle $\theta$, (c) yaw angle $\phi$, and (d) hydrodynamic force components and lubrication force. The control parameters are $I^* = 5.97 \times 10^{-3}$ and *Re* = 255.

## 5.4 Separation after a single collision

The falling pattern also depends on the shape of the disks. At the Reynolds number of *Re* = 320, by increasing the disk aspect ratio *AR* from 0.1 to 0.4 ($I^*$ from $5.97 \times 10^{-3}$ to $2.86 \times 10^{-2}$), the two disks collide once and thereafter separate, without periodic oscillation of the horizontal translation. As shown in Figure 15a, the two disks initially fall along the vertical central line ($Z^* = 0$) until they collide at about $Y^* = 18$. The trajectories of the two disks deviate from the central line after the collision. At the early stage, the two disks rotate to have the major axes aligned vertically with $\theta$ close to 0°, as shown in Figure 15b. The collision at $Y^* = 18$ causes stronger rotation of the two particles, which is thereafter reduced as the two disks settle separately. The nontrivial magnitudes of the yaw angle $\phi$ are observed after the collision, as shown in Figure 15c, illustrating the three-dimensional rotation, rather than the planar motion, of the disks. As shown in the insert of Figure 15d, non-zero lubrication force $F_L^*$ is obtained at about $Y^* = 18$ as the collision occurs. After the collision, the contributions of the hydrodynamic forces for the disks to contact or separate are significant reduced, because the angles between the hydrodynamic force vectors $\boldsymbol{F}_H^i$ and the branch vector $\boldsymbol{r}_{ji}$ increase, resulting in the reduced projections of the hydrodynamic forces in the branch vector direction. This situation is associated with the side-by-side falling of the two disks. In addition, after the collision the initially trailing disk (disk 1) surpasses the other one (disk 2) and becomes the new leading disk. As a result, the hydrodynamic force component $F_H^{*1}$ changes its sign from positive to negative and $F_H^{*2}$ changes its sign from negative to positive. The drafting, kissing, and tumbling (DKT) behaviors, which frequently occur in the settling of a pair of spheres at high Reynolds numbers, are also observed in this sedimentation with the two disks of $I^* = 2.86 \times 10^{-2}$ and $Re = 320$. The above falling pattern with $AR = 0.4$ and $Re = 320$ is similar to the pattern with $AR = 1/3$ and $Re = 255$ and 285 observed in the experiments by Brosse & Ern (2014) and Ern & Brosse (2014), in which the two thick disks ($AR = 1/3$) separated after contact and fell side by side.

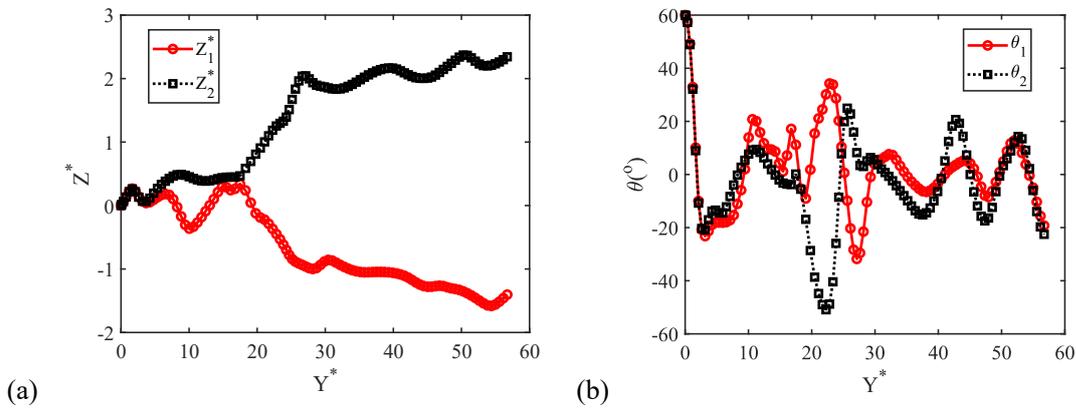

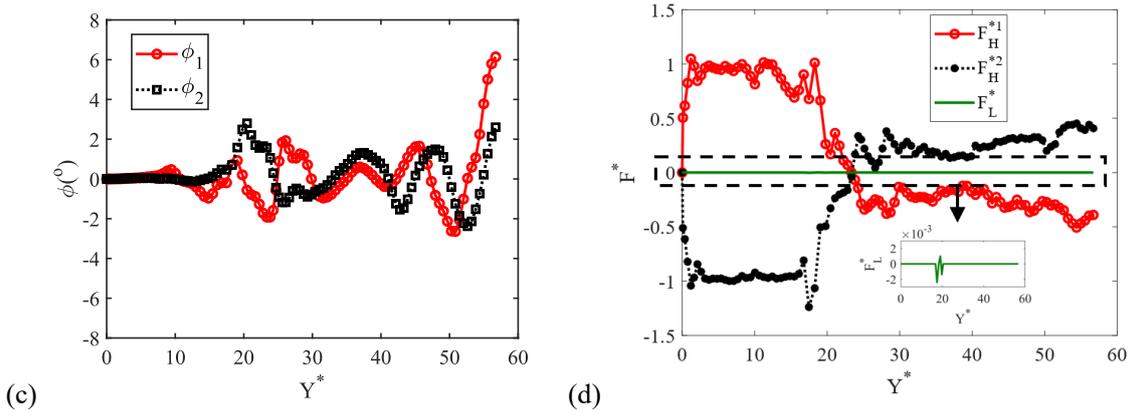

Figure 15: Evolution of (a) horizontal displacements in the *z* direction, (b) pitch angles $\theta$, (c) yaw angles $\phi$, and (d) hydrodynamic force components and lubrication force obtained from the simulation. The control parameters are $I^* = 2.86 \times 10^{-2}$ and $Re = 320$.

### 5.5 Tumbling without disk-disk contact

For the disks of a large dimensionless moment of inertia $I^* = 1.37 \times 10^{-1}$ (aspect ratio $AR = 1$) at a very large Reynolds number $Re = 1000$, the disks fall in a tumbling pattern with nearly random orientations and no disk-disk contact occurs, as shown in Figure 16. The motions of the disks are not constrained in the *y-z* plane, and displacements in the *x*-direction are also observed. As the disks remain separated, high- and low-pressure regions, which are created in the front of and behind each individual disk (Figure 17a), are much smaller than those formed near the contacting pair of disks (Figures 10a and 13a). Associated with the complex motion of the disks, the structures of the vortices induced by the disks are very chaotic, as shown in Figure 17b.

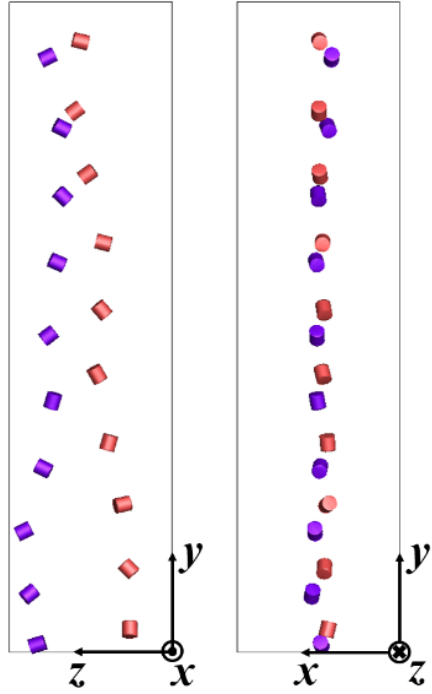

Figure 16: Sequential snapshots of the falling disks superimposed in the same plot of domain obtained from the simulation. The two pictures are obtained from different view angles. The control parameters are $I^* = 1.37 \times 10^{-1}$ and $Re = 1000$.

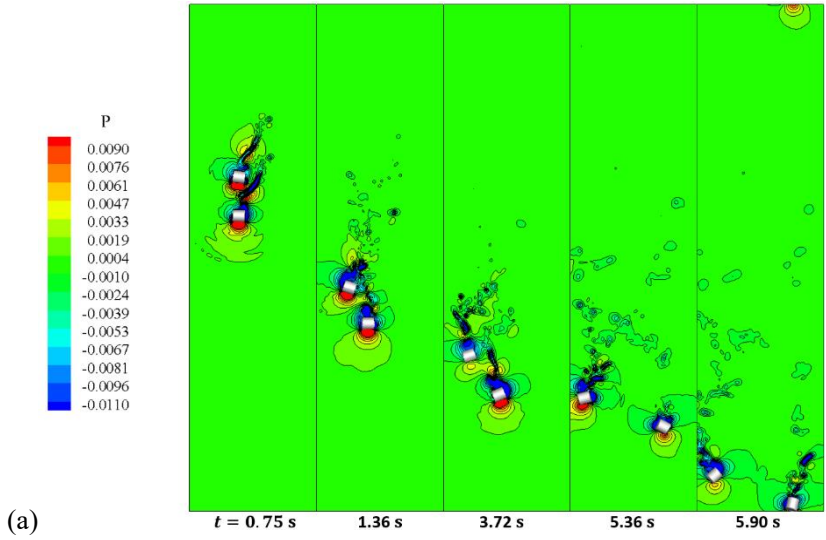

(a)

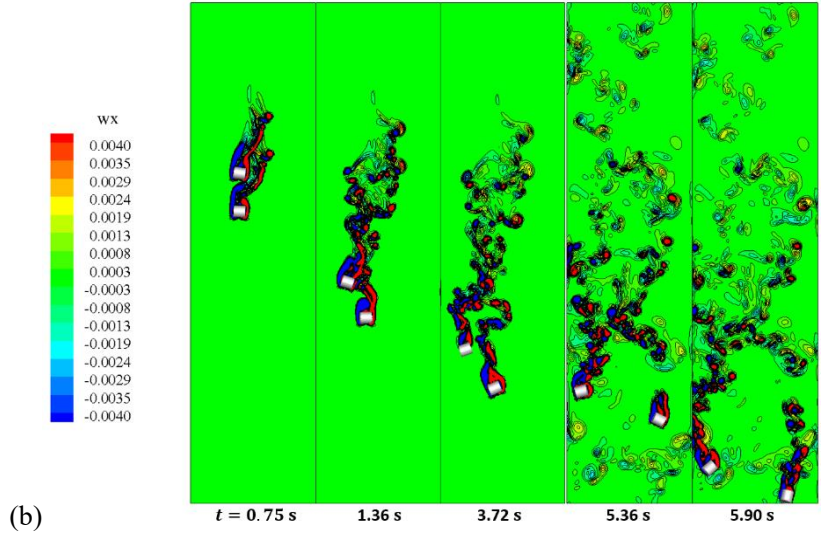

(b)   t = 0.75 s   1.36 s   3.72 s   5.36 s   5.90 s

Figure 17: Contours of (a) fluid pressure $P$ and (b) fluid vorticity in the $x$ direction $w_x$ at various time instants $t$ in the sedimentation of two disks obtained from the simulation. The views are in the $y$-$z$ plane. The control parameters are $I^* = 1.37 \times 10^{-1}$ and $Re = 1000$. The movies of the simulation and physical experiment are provided in Supplementary Materials.

For the settling pattern of segregation without contact, the two disks deviate from the vertical central line at the very beginning and follow inclined paths thereafter, as shown in Figure 18a. The pitch angle $\theta$ varies in a wide range of $-90°$ to $90°$ (Figure 18b), due to the tumbling of the disks. The yaw angles $\phi$ oscillate between $-20°$ and $20°$ (Figure 18c), and thus the two disks undergo three-dimensional rotations. Consistent with the observation of no contact, the lubrication force $F_L^*$ remains zero in the whole settling process, as shown in Figure 18d. In addition, the magnitudes of the hydrodynamic force components, which contribute to the contact or separation of the two disks, decrease as the separated disks settle side-by-side.

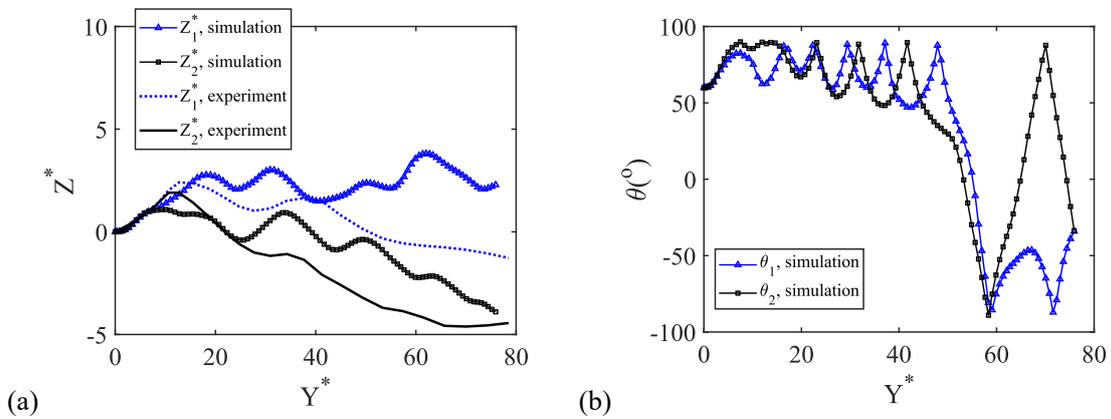

(a)   (b)

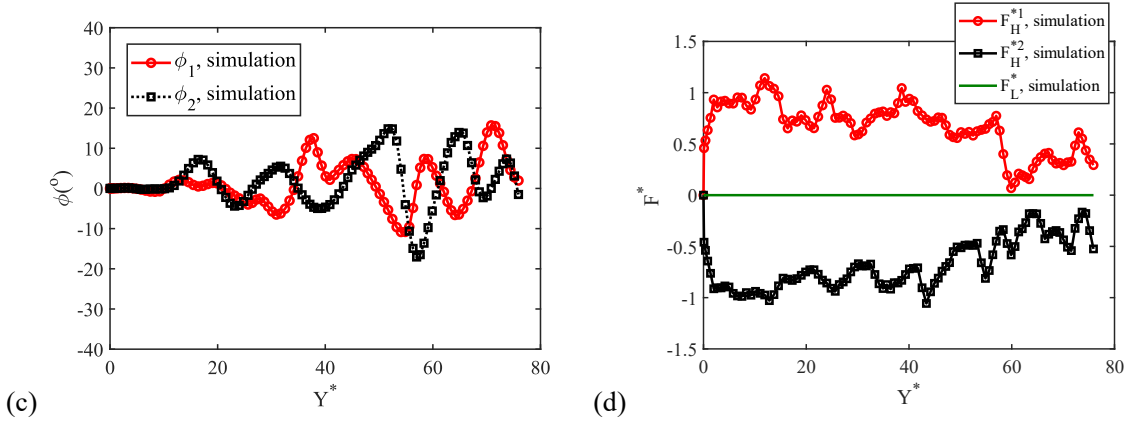

Figure 18: Evolution of (a) horizontal displacements in the $z$ direction, (b) pitch angles $\theta$, (c) yaw angles $\phi$, and (d) hydrodynamic force components and lubrication force. The control parameters are $I^* = 1.37 \times 10^{-1}$ and $Re = 1000$.

The evolution of the ratio of the angular-velocity component in the axis-of-symmetry direction, $\omega_l$, to the angular-velocity component in a diameter direction, $\omega_d$, is plotted for each disk in Figure 19. For the settling in a tumbling mode with $I^* = 1.37 \times 10^{-1}$ and $Re = 1000$, as shown in Figure 19a, the angular-velocity component $\omega_l$ is significant compared to the component $\omega_d$, indicating the remarkable disk rotation about the axis of symmetry. While in the periodic swinging sedimentation with $I^* = 5.97 \times 10^{-3}$ and $Re = 255$, as shown in Figure 19b, the angular velocity component $\omega_l$ is negligible compared to the component $\omega_d$ for most of the settling process. Thus, the rotation about a diameter is more dominant than the rotation about the axis of symmetry for the disks in the periodic swinging pattern, and the relative importance of the rotation about the axis of symmetry compared to that about a diameter increases in the tumbling pattern. The flatter disks with a smaller $I^*$ have a larger moment-of-inertia ratio $I_l/I_d$, leading to greater resistance to the rotation about the axis of symmetry. Thus, a larger moment-of-inertia ratio $I_l/I_d$ contributes to the more dominant rotation about a diameter than about the axis of symmetry for the disks in the sedimentation process.

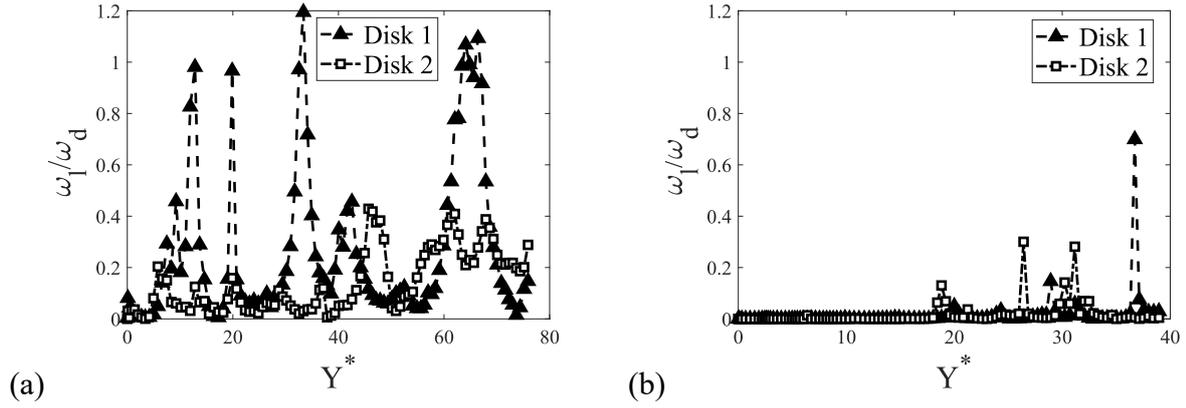

Figure 19: Evolution of ratios of angular velocity components obtained from the simulations for the control parameters: (a) $I_l/I_d = 0.857$, $I^* = 1.37 \times 10^{-1}$, and $Re = 1000$; (b) $I_l/I_d = 1.97$, $I^* = 5.97 \times 10^{-3}$, and $Re = 255$.

## 5.6 Phase diagrams of falling patterns

The present results and the previous results (Brosse & Ern, 2011; 2014) show that the falling styles of the two identical disks in a viscous fluid are determined jointly by the dimensionless moment of inertia of the disks $I^*$ and the terminal Reynolds number of the leading disk $Re$. Thus, based on $I^*$ and $Re$, a phase diagram is generated in Figure 20a to classify the two-disk sedimentation into ten patterns: 1. Steady falling with enduring disk-disk contact; 2. Periodic swinging with intermittent disk-disk contacts; 3. Three-dimensional oscillating with intermittent contacts; 4. Separation after a single collision and steady falling with major axes of the disks aligned vertically; 5. Separation after a single collision and chaotic three-dimensional oscillating with the major axes aligned almost vertically; 6. Falling without disk-disk contact and three-dimensional oscillating with the major axes aligned almost vertically; 7. Separation after a single collision and steady falling with major axes aligned horizontally; 8. Separation after a single collision and steady falling with no preferential alignment of the major axes; 9. Falling without disk-disk contact and three-dimensional oscillating with the major axes aligned almost horizontally; 10. Tumbling without disk-disk contact. For the flatter disks with a small value of $I^* = 5.97 \times 10^{-3}$, steady falling with enduring contact occurs when $Re < 100$, while 2D periodic swinging with intermittent contacts takes place when $100 < Re < 270$, and the 2D periodic swinging transits to a 3D periodic oscillation when $Re > 270$. The disks with larger values of $I^*$ settle in a Drafting-Kissing-Tumbling (DKT) mode when $10 < Re < 100$, and the disks fall steadily after the separation. When $Re > 200$, the increase in $I^*$ leads to single or no collision and the chaotic 3D oscillating motions of the disks. A detailed description of the major features of the two-disk falling patterns is provided in Appendix C.

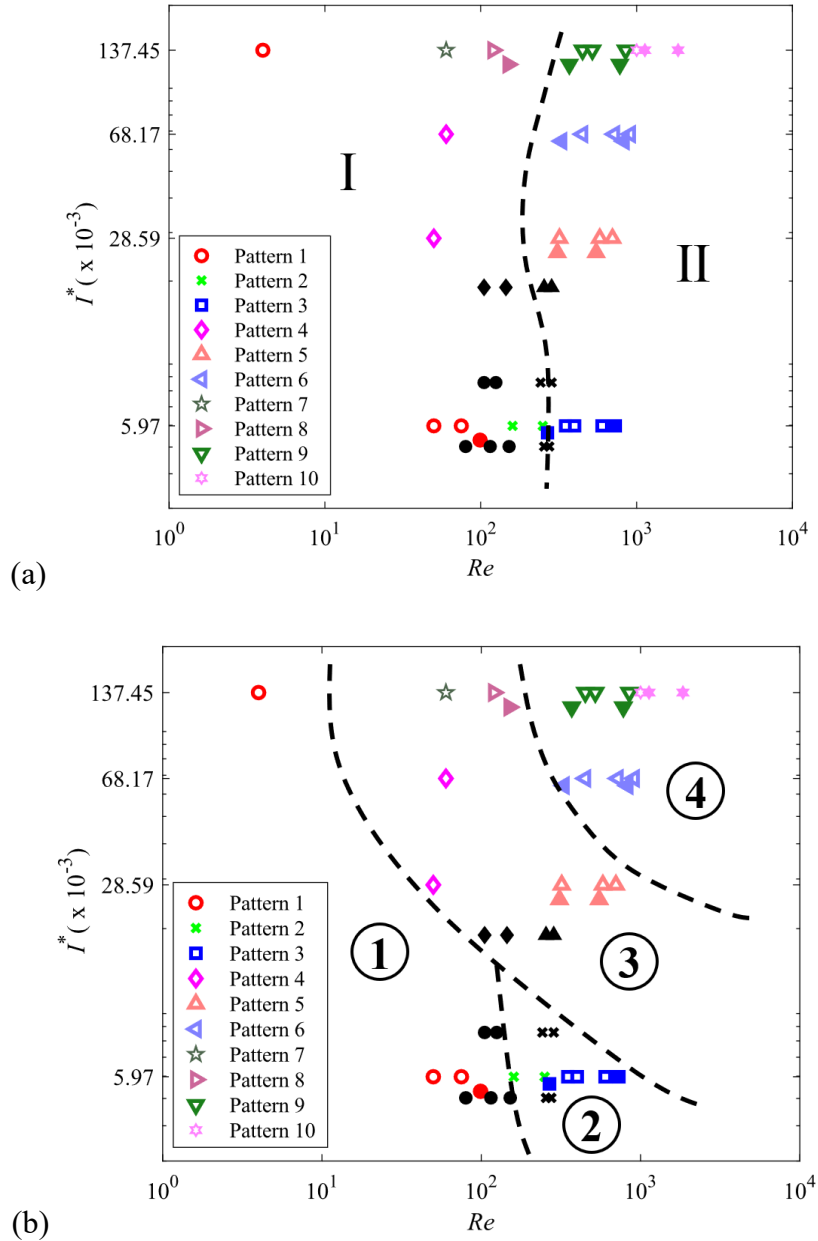

Figure 20: Phase diagrams of two-disk falling patterns: (a) planar motion (regime I) versus three-dimensional motion (regime II) and (b) regimes of various disk-disk contact types: ①enduring contact, ②intermittent and multiple contacts, ③a single contact, and ④no contact. Open marks represent the present simulation results and filled marks are the present experimental results. The black filled marks represent the experimental data from Brosse & Ern (2011; 2014).

As shown in Figure 20a, the phase diagram is partitioned by a black dash curve into regimes I and II, which are associated with the planar motion and three-dimensional (3D) motion, respectively, of the disks. Thus, the planar or 3D motion is determined primarily by the Reynolds number $Re$ regardless of the aspect ratio $AR$ or dimensionless moment of inertia $I^*$ of the disks, because the stronger turbulent flows enhance the 3D rotation of the disks (see Secs. 5.4 and 5.5). This conclusion is valid for the symmetric disk-like objects. However, it was observed that the asymmetric rigid fibers settled in 3D helical paths at very low $Re$ (<1) in the previous experiments (Tozzi et al., 2011). As a result, the change in the object shape from symmetry to asymmetry can have a significant impact on the planar or 3D motion of the objects.

By comparing the phase diagram of two-disk falling patterns (Figure 20a) with that of a single disk falling (Figure A2 in Appendix A1), the 3D falling patterns of the two disks are related to the 3D motion of a separated disk in the sedimentation. As shown in Figure 20a, the 3D falling of the two disks occurs at about $Re > 3 \times 10^2$. For a separated disk, as shown in Figure A2 in Appendix A1, when $Re > 3 \times 10^2$, the thicker disks of $I^* > 4 \times 10^{-2}$ settle in a tumbling mode, the moderately-thick disks of $10^{-2} < I^* < 4 \times 10^{-2}$ in a chaotic mode, and the thinner disks of $I^* < 10^{-2}$ in a periodic mode. The tumbling and chaotic modes are essentially characterized as complex 3D motions. The periodic oscillations of a separated thinner disk at high $Re$ exhibited a planar Zigzag pattern (Zhong et al., 2011; Lee et al., 2013) or a 3D hula-hoop pattern (Auguste et al., 2013). Thus, it is interesting to note that the two thinner disks at a high $Re$ (> $3 \times 10^2$), at which a separated same disk falls in a planar Zigzag pattern, can settle in 3D motion due to the hydrodynamic interaction between them.

According to the types of disk-disk contacts in the sedimentation process, the phase diagram can be also classified into four distinctive regimes: ①enduring contact, ②intermittent and multiple contacts, ③a single contact, and ④no contact, as shown in Figure 20b. Therefore, the contacts are affected by a combination of $I^*$ and $Re$, and the probability and duration of the disk-disk contact are increased by reducing $I^*$ and $Re$. It is noted that the phase diagrams shown in Figure 20 are valid for the disks with the initial inclination angles $\theta_0$ between 30° and 80°, and the disks with zero initial inclination angles may exhibit different falling behaviors according to the studies on the effects of initial conditions in Section 5.1.

The evolutions of the distance $D$ between the mass centers of two disks, scaled by the equivalent volume sphere diameter $d_{eq}$, for various falling patterns are shown in Figure 21, in which $D^* = D/d_{eq}$. The evolutions of the corresponding scaled lubrication forces $F_L^*$ are plotted in Figure 22. For the enduring-contact falling (Pattern 1), the scaled distance $D^*$ maintains as about one in the contacting process and $F_L^*$ is present as the contact persists. For the intermittent and multiple contacts (Patterns 2 and 3), $D^*$ oscillates around one and $F_L^*$ appears occasionally. For the single-contact fallings (Patterns 4, 5, 7, and 8), $D^*$ initially decreases and then increases. The non-zero value of $F_L^*$ is invoked only when the two disks move very close to each other (corresponding to the small $D^*$), and $F_L^*$ disappears when $D^*$ increases. The falling style in this single-contact regime is analogous to the DKT of a pair of spheres settling in a fluid. For the no-contact fallings (Patterns 6, 9, and 10), $D^*$ generally increases from the beginning and $F_L^*$ vanishes in the whole falling process.

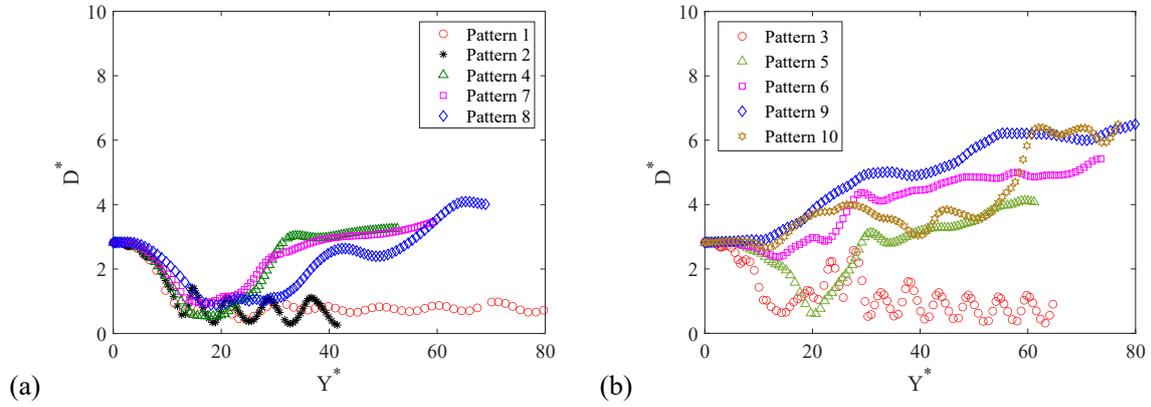

Figure 21: Evolutions of the distance between the mass centers of two disks, scaled by the equivalent volume sphere diameter $d_{eq}$, for various falling patterns with (a) planar motion and (b) three-dimensional motion of the disks. The data are obtained from the simulations.

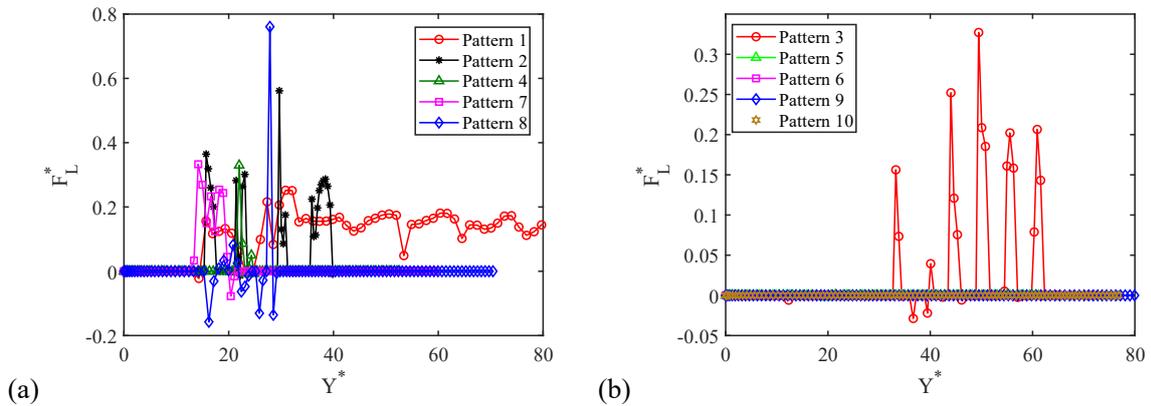

Figure 22: Evolutions of the scaled lubrication force between the two disks $F_L^*$ for various falling patterns with (a) planar motion and (b) three-dimensional motion of the disks. The data are obtained from the simulations.

## 6. Conclusions

Solid-surface-resolved Direct Numerical Simulations (DNS) of a pair of disks falling in a fluid are performed using a coupled approach of Lattice Boltzmann Method (LBM) and cylindrical particle Discrete Element Method (DEM). The simulation results are generally consistent with the present and previous experimental observations (Brosse & Ern, 2011; 2014). Based on the simulations, translational and rotational displacements of the disks are presented to illustrate the sedimentation patterns, the contributions of hydrodynamic forces to collision or separation of the disks are discussed, and the evolution of the lubrication force is analyzed to understand the disk-disk contact scenarios.

It is observed that the settling behaviors are determined by the combination of a dimensionless moment of inertia of the disks $I^*$ and a disk Reynolds number $Re$. For the flatter disks with a small value of $I^* = 5.97 \times 10^{-3}$, steady falling with enduring contact occurs when $Re < 100$, while 2D periodic swinging with intermittent contacts takes place when $100 < Re < 270$, and the 2D periodic swinging transits to a 3D periodic swinging when $Re > 270$. The disks with larger values of $I^*$ fall in a Drafting-Kissing-Tumbling (DKT) mode when $10 < Re < 100$. When $Re > 200$, the increase in $I^*$ leads to single or no collision and the chaotic 3D oscillating motions of the disks. Based on $I^*$ and $Re$, a planar phase diagram is created to classify the two-disk falling into ten distinctive patterns. The planar motion or three-dimensional motion of disks is determined primarily by $Re$. It is believed that the strong disturbance flows at a high $Re$ contribute to the chaotic 3D rotation of disks. The chance for the two disks to contact is increased when $I^*$ and $Re$ are reduced.

This work presents a primary study on the hydrodynamic interactions of two non-spherical particles in the sedimentations. Future work may be conducted to understand the effects of solid wall boundaries on the sedimentation of a pair of non-spherical particles and the settling behaviors of multiple non-spherical particles.

**Declaration of Interests**

**Acknowledgement**

The National Science Foundation of China (NOs. 91852205, 11872333, 12132015, 91752117, and 11988102), the Zhejiang Provincial Natural Science Foundation of China (NO. LR19A020001), and the Fundamental Research Funds for the Central Universities (NO. 2021FZZX001-11) are acknowledged for their financial support.

The authors also express their gratitude to (i) Dr. Zhiqiang Dong for the helpful discussion in the numerical code development and (ii) Dr. Guichao Wang for providing the MATLAB code to extract data from videos taken in the physical experiments. At the time, both Dr. Dong and Dr. Wang worked at the Southern University of Science and Technology, Shenzhen, China.

**Appendix A: Validation of the coupled scheme of Lattice Boltzmann Method (LBM) and cylindrical particle Discrete Element Method (DEM)**

Four sets of simulations have been performed using the coupled LBM-DEM code developed in this work. The present simulation results are compared with the existing experimental and numerical results for the code validation.

**A1. Falling modes for a single disk**

Using the present LBM-DEM method, sedimentations of a single disk of $AR = 0.1$ and $I^* = 5.97 \times 10^{-3}$ at various particle terminal Reynolds numbers are simulated. A three-dimensional rectangular computational domain of dimensions $L_x = L_z = 300$ (lattice unit or LU) and $L_y = 1200$ (LU) is created, as shown in Figure 5a. The disk of an equivalent volume sphere diameter of $d_{eq} = 31.88$ (LU) is placed at the position of the coordinates ($L_x/2$, $L_y - d_1$, $L_z/2$). The clearance to the top boundary is specified as $d_1 = 6.27 d_{eq}$. Here, the Reynold number is defined as,

$$Re = \frac{u_c d_c}{\nu_f}, \tag{A1}$$

in which $u_c$ and $d_c$ are the terminal vertical velocity and diameter, respectively, of the disk, and $\nu_f$ is the kinematic viscosity of the fluid. The disk is released with the initial angle of 60º between the axis of the disk and the vertical direction, and it falls under the gravitational force. The sequential snapshots of the falling disk at different $Re$ are shown in Figure A1. At a low Reynolds number ($Re = 55$), the disk eventually achieves a steady falling mode with the axis of the disk aligned vertically. At a higher Reynolds number ($Re = 120$), periodic movement in the horizontal direction and periodic rotation of the disk are observed. The further increases in

Reynolds number ($Re$ = 180 and 350) lead to an increase in the amplitudes of the periodic motion.

Based on the dimensionless moment of inertia $I^*$ and particle Reynolds number $Re$, Field et al. (1997) classified four distinctive falling modes for a single disk in a viscous fluid: steady falling, periodic, chaotic, and tumbling, as shown in Figure A2. The falling modes obtained from the present simulations are consistent with the classification by Field et al. (1997) (see Figure A2).

Using the experimental set-up described in Section 4, the sedimentation of a single acrylic disk in a solution of glycerol at $Re$ = 180 is conducted with an initial inclination angle of $\theta_0$ = 60°. The disk has a density of 1.2 g/cm³, a diameter of $d_c$ = 20 mm, a thickness of $l_c$ = 2 mm, and thus $AR$ = 0.1. A comparison between the simulation and experimental results is made for the horizontal displacements $Z^*$ and pitch angles $\theta$ in Figures A3a and A3b, respectively. The experimental data from three runs under the same conditions are presented. Similar sinusoidal oscillations are obtained in the simulation and experiments. The differences between the simulation and experiments in the oscillatory magnitudes and frequencies are about 15%, which is comparable to the extent of the difference between two experimental runs. In the experiment, a disk was held and released by a clamp, which might affect the initial falling of the disk by interacting with the surrounding fluid. The clamp did not exist in the numerical simulation. Thus, the initial conditions of the disk falling were not exactly same between the experiment and simulation, which could cause the discrepancy in the comparison.

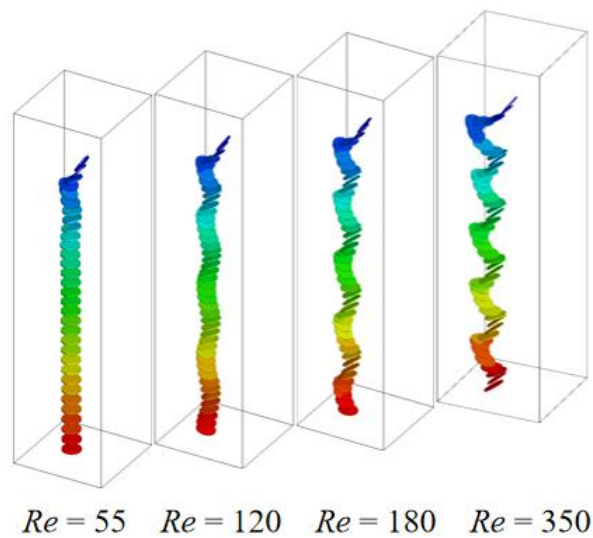

$Re = 55 \quad Re = 120 \quad Re = 180 \quad Re = 350$

Figure A1: The sequential snapshots of the falling disk at different $Re$.

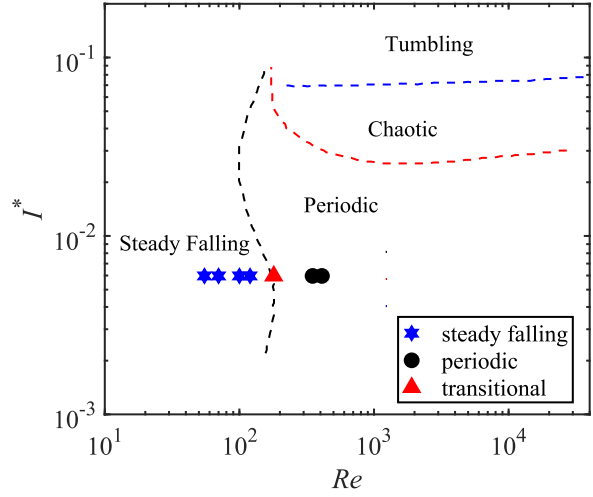

Figure A2: Phase diagram of the typical modes of a disk falling in a viscous fluid.

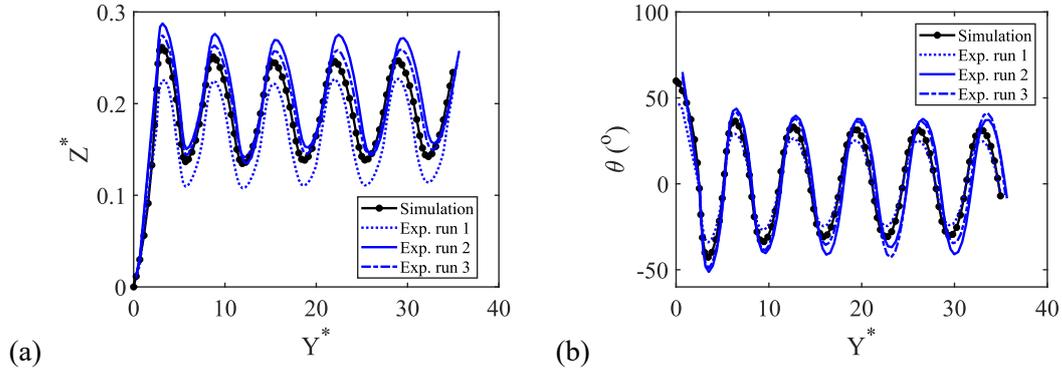

Figure A3: Evolution of (a) horizontal displacements in the $z$ direction $Z^*$ and (b) pitch angle $\theta$ of the settling disk for the control parameters $I^* = 5.97 \times 10^{-3}$ and $Re = 180$.

**A2. Drag coefficient versus particle terminal Reynolds number**

Clift et al. (1978) experimentally determined the correlation between drag coefficient $C_D$ of a disk falling in a fluid and particle terminal Reynolds number $Re$, as shown in Figure A4. The drag coefficient is defined as,

$$C_D = \frac{F_D}{\frac{1}{2}\rho_f u_t^2 \frac{\pi}{4} d_{eq}^2}, \tag{A2}$$

in which, $F_D$ is the drag force on the disk (i.e. the hydrodynamic force in the vertical direction), $u_t$ is the terminal falling velocity of the disk, and $d_{eq}$ is the equivalent volume sphere diameter of the disk. Two regimes can be distinguished: (i) For $Re$ < 100, the disk eventually achieves a steady falling mode, and thus the drag coefficient exhibits a quick decrease with increasing $Re$.

(ii) For *Re* > 100, secondary motion of the disk (horizontal movement and rotation) occurs, associated with wake shedding at the edge of the disk, leading to the drag coefficient insensitive to the Reynolds number *Re*.

Sedimentations of a single disk ($AR = 0.1$ and $I^* = 5.97 \times 10^{-3}$) at various *Re* are simulated using the present coupled LBM-DEM method, and the computational set-up is the same to that in the Appendix A1. The drag coefficients obtained from the simulations are also plotted in Figure A3. A comparison shows that the present simulation results are in very good agreement with the correlation by Clift et al. (1978) in both *Re*-sensitive and *Re*-insensitive regimes.

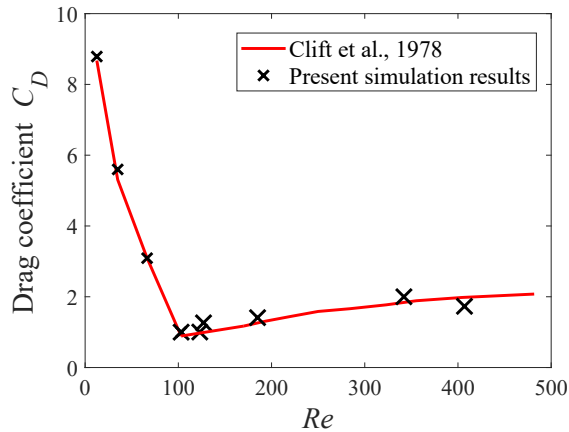

Figure A4: Drag coefficient versus *Re* for a falling disk of $AR = 0.1$ and $I^* = 5.97 \times 10^{-3}$ in a fluid.

## A3. Lift and torque coefficients for a cylindrical particle

Fluid flows around a cylinder fixed at a specified position and orientation are simulated using the present coupled LBM-DEM approach. As illustrated in Figure A5a, a rectangular domain of $x \times y \times z$ dimensions of $200 \times 900 \times 600$ (lattice unit or LU) is created, and a cylinder, which has a diameter $d_c = 12$ LU, a length $l_c = 60$ LU, and thus an aspect ratio $AR = l_c/d_c = 5$, is fixed at the center of the domain. The effect of the cylinder orientation is examined by adjusting the attack angle $\alpha$ in the *x-y* plane, as shown in Figure A5b. The fluid flows along the *y* direction from the left-hand side to the right-hand side. A constant fluid velocity *U* is specified on the inlet boundary. The outlet boundary and the boundaries in the *y* and *z* directions are the fully developed flows with zero gradients of fluid velocities and pressures. In the simulations, the cylinder Reynolds number is specified as $Re = \frac{U d_{eq}}{\nu_f} = 300$, in which $d_{eq}$ is the equivalent volume sphere diameter of the cylinder.

With an attack angle $\alpha$, the lift force $F_L$ perpendicular to the streamwise $y$ direction can be calculated. Thus, the lift coefficient $C_L$ can be obtained as

$$C_L = \frac{F_L}{\frac{1}{2}\rho_f U^2 \frac{\pi}{4} d_{eq}^2} . \tag{A3}$$

The lift coefficient $C_L$ is plotted as a function of the attack angle $\alpha$ in Figure A6a, and good agreement is obtained between the present simulation results and the previous DNS simulation results of an ellipsoid with the same aspect ratio of $AR = 5$ by Zastawny et al. (2012). Also, the present simulation results follow the prediction by the models proposed by Kharrouba et al. (2021), which are available for the attack angles $\alpha \leq 30°$. In the article by Kharrouba et al. (2021), the Reynolds number is defined as $Re^* = \frac{U d_c}{\nu_f}$ and thus the present $Re$ and the previous $Re^*$ follow the correlation

$$Re^* = Re \left(\frac{2}{3AR}\right)^{\frac{1}{3}} . \tag{A4}$$

The parallel and perpendicular coefficients are defined as

$$C_\parallel^* = \frac{F_\parallel}{\frac{1}{2}\rho_f U^2 l_c d_c} , \tag{A5}$$

$$C_\perp^* = \frac{F_\perp}{\frac{1}{2}\rho_f U^2 l_c d_c} , \tag{A6}$$

in which $F_\parallel$ and $F_\perp$ represent the hydrodynamic force components parallel and perpendicular, respectively, to the symmetric axis of the cylindrical object. The lift force can be written as

$$F_L = F_\perp \cos\alpha - F_\parallel \sin\alpha . \tag{A7}$$

From the Eqs.(A3-A6), we can obtain

$$C_L = \frac{4}{\pi} \left(\frac{2}{3}\right)^{\frac{2}{3}} AR^{\frac{1}{3}} \left(C_\perp^* \cos\alpha - C_\parallel^* \sin\alpha\right) . \tag{A8}$$

The mathematical expressions of $C_\perp^*$ and $C_\parallel^*$ as functions of $AR$, $Re^*$, and $\alpha$ can be found in Kharrouba et al. (2021).

Pitching torque (in the $z$ direction) $T_P$ exerted on the cylinder can be also obtained from the simulations, and a torque coefficient $C_T$ is defined as

$$C_T = \frac{T_P}{\frac{1}{2}\rho_f U^2 \frac{\pi}{8} d_{eq}^3} . \tag{A9}$$

Figure A6b shows the torque coefficient as a function of the attack angle. The present simulation results are consistent with the previous simulation results (Zastawny et al., 2012) and the prediction of the torque coefficient model (Kharrouba et al., 2021). In the work by Pierson et al. (2019), the torque coefficient $C_T^*$ is defined as

$$C_T^* = \frac{T_P}{\frac{1}{2}\rho_f U^2 l_c^2 d_c}. \tag{A10}$$

Thus, the present torque coefficient $C_T$ is related to $C_T^*$ as

$$C_T = \frac{16AR}{3\pi} C_T^* . \tag{A11}$$

The mathematical model of $C_T^*$ as a function of $AR$, $Re^*$, and $\alpha$ is provided in Kharrouba et al. (2021).

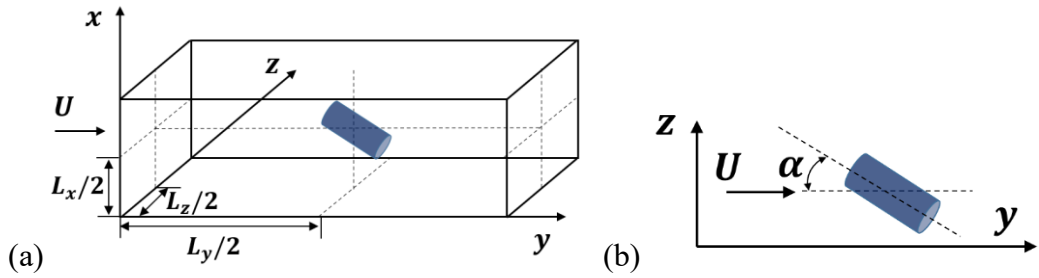

Figure A5: (a) A numerical domain of a fluid flowing around a fixed cylinder, and (b) an illustration of attack angle α: the angle between the major axis of the cylinder and the streamwise direction *y* in the *y-z* plane.

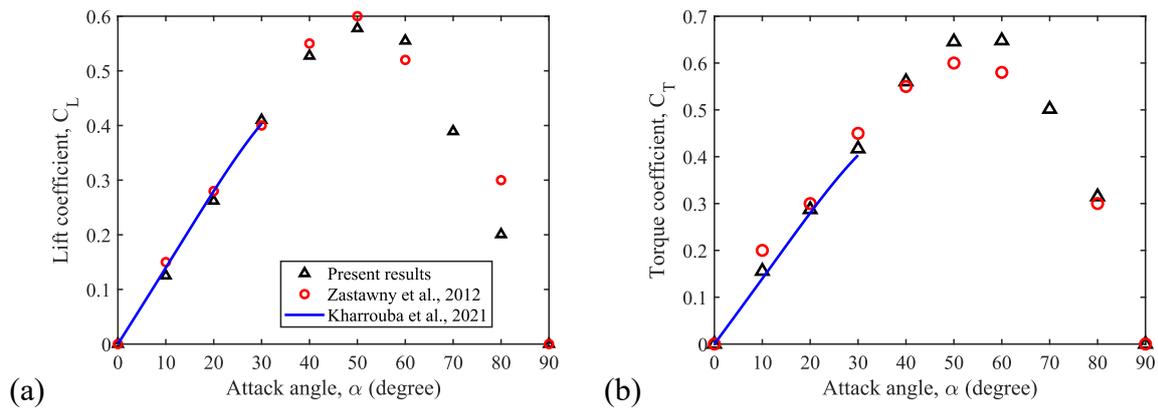

Figure A6: (a) Lift coefficient and (b) torque coefficient of a single cylinder as a function of attack angle α.

## A4. Two-disk falling in tandem

In a simulation of two identical disks falling in tandem, the disks have an aspect ratio of $AR$ = 0.1 and a diameter of $d_c$ = 9 mm. The densities of fluid and disks are $\rho_f$ = 1010 kg/m³ and $\rho_s$ = 1020 kg/m³, respectively. The Reynolds number of the leading disk is $Re = \frac{u_l d_c}{v_f}$ =115. These parameters in the present simulation are the same to those in an experimental case by Brosse & Ern (2011). In the experiment (Brosse & Ern, 2011), the disks were released at separate times through a 20-cm long tube. The velocities and orientation of the disks and the vertical distance between them were unclear when they moved out of the tube. Thus, it is impossible to ensure the same initial conditions for the previous experiment and the present simulation. To produce the similar process of two disk falling in tandem, in the present simulation two disks are released simultaneously with a small initial inclination angle of $\theta_0$ = 5° and an initial distance of $d_2$ = 2.282 $d_{eq}$ (see Figure 5). It can be seen that the trailing disk catches up with the leading one, and the two disks fall together steadily in a Y-configuration. This falling pattern is similar to the experimental observation by Brosse & Ern (2011) with same set of parameters ($AR$ = 0.1 and $Re$ = 115). The quantitative comparison between the experimental and simulation results is made in Figure A7. The similar periodic oscillation patterns of the horizontal displacement $z/d_c$ and inclination angle $\theta$ are obtained between the experiment and simulation. However, larger periods and magnitudes of the oscillations are observed in the simulation than in the experiment. The discrepancies may be attributed to the differences in the initial conditions between the experiment and simulation. In addition, a small disturbance in the fluid can cause a notable change in the dynamics of disks, causing the differences even between two independent experimental runs. It is very hard to ensure completely identical conditions between the experiment and simulation.

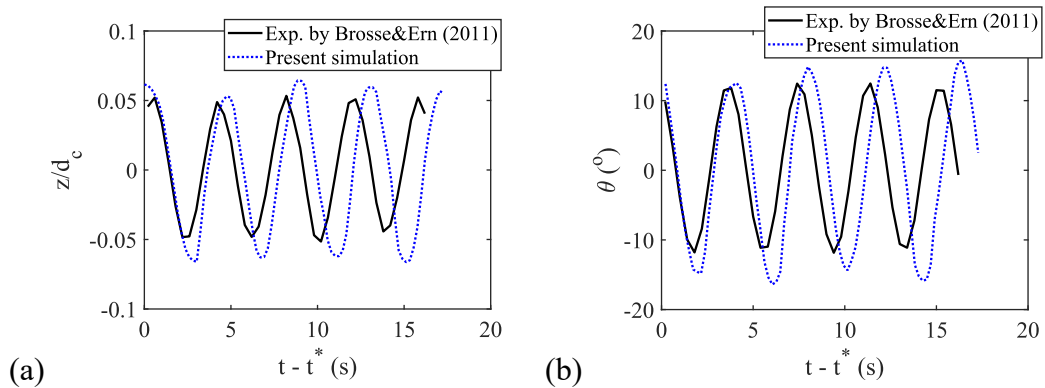

Figure A7. Time evolution of (a) the normalized horizontal displacement $z/d_c$ and (b) the inclination angle $\theta$ at the steady falling state. The time instant $t^*$ is 40 s for the experiment, and $t^*$ is 21.3 s for the simulation to ensure that the comparison starts from the same phase position of the periodic oscillations. The controlling parameters are $AR = 0.1$ and $Re = 115$.

**Appendix B: Conversion of the lattice units used in the Lattice Boltzmann Method (LBM) to the international system of units (SI)**

The conversion factor of length is expressed as,

$$C_H = \frac{\tilde{D}}{D}, \tag{B1}$$

in which $\tilde{D}$ and $D$ represent the cylinder diameters in physical unit and lattice unit, respectively. To ensure the equality of Reynolds numbers between the physical unit system and lattice unit system, the conversion factor of time is derived as,

$$C_t = \frac{C_H^2}{C_v}, \tag{B2}$$

where $C_v$ is the conversion factor of kinematic viscosity and has the form,

$$C_v = \frac{\tilde{v}}{v}, \tag{B3}$$

and $\tilde{v}$ and $v$ are the kinematic viscosities in the physical unit and lattice unit, respectively. Thus, the conversion factor of velocity is obtained as,

$$C_u = \frac{C_H}{C_t}. \tag{B4}$$

The velocity $\tilde{u}$ and time $\tilde{t}$ in the physical units can be determined using the corresponding quantities $u$ and $t$ in the lattice units,

$$\tilde{u} = u \cdot C_u, \tag{B5}$$

and

$$\tilde{t} = t \cdot C_t. \tag{B6}$$

**Appendix C: Features of the two-disk falling patterns**

The major features of the ten falling patterns are described in Table C1.

| Pattern No. | Description | Disk dynamics | 2D/3D motion | Type of contacts | Orientation of disks (direction of major axis of the disk) |
|---|---|---|---|---|---|
| 1 | Steady falling with enduring disk-disk contact | Steady | 2D | Enduring | Vertical |
| 2 | Periodic swinging with intermittent disk-disk contacts | Periodic oscillating | Transitional | Multiple | Oscillating about vertical axis |
| 3 | Three-dimensional oscillating with intermittent contacts | Periodic oscillating | 3D | Multiple | Oscillating about vertical axis |
| 4 | Separation after a single collision and steady falling with major axes of the disks aligned vertically | Steady | 2D | Single | Vertical |
| 5 | Separation after a single collision and chaotic three-dimensional oscillating with the major axes aligned almost vertically | Chaotic oscillating | 3D | Single | Oscillating about vertical axis |
| 6 | Falling without disk-disk contact and three-dimensional oscillating with the major axes aligned almost vertically | Chaotic oscillating | 3D | No contact | Oscillating about vertical axis |
| 7 | Separation after a single collision and steady falling with major axes aligned horizontally | Steady | 2D | Single | Horizontal |
| 8 | Separation after a single collision and steady falling with no preferential alignment of the major axes | Steady | 2D | Single | Randomly horizontal or vertical |
| 9 | Falling without disk-disk contact and three-dimensional oscillating with the major axes aligned almost horizontally | Chaotic oscillating | 3D | No contact | Oscillating about horizontal axis |
| 10 | Tumbling without disk-disk contact | Tumbling | 3D | No contact | Randomly oscillating and rotating |

Table C1: Features of the two-disk falling patterns.

*References*

Aidun, C.K. & Ding, E.-J. 2003 Dynamics of particle sedimentation in a vertical channel: Period-doubling bifurcation and chaotic state. *Physics of Fluids* 15, 1612-1621.

Auguste, F., Magnaudet, J. & Fabre, D. 2013 Falling styles of disks. *Journal of Fluid Mechanics* 719, 388-405.

Ardekani, M.N., Costa, P., Breugem, W.-P. & Brandt, L. 2016 Numerical study of the sedimentation of spheroidal particles. *International Journal of Multiphase Flow* 87: 16–34

Bouzidi, M., Firdaouss, M. & Lallemand, P. 2001 Momentum transfer of a Boltzmann-lattice fluid with boundaries. *Physics of Fluids* 13, 3452-3459.

Brosse, N. & Ern, P. 2011 Paths of stable configurations resulting from the interaction of two disks falling in tandem. *Journal of Fluids and Structures* 27, 817–823.

Brosse, N. & Ern, P. 2014 Interaction of two axisymmetric bodies falling in tandem at moderate Reynolds numbers. *Journal of Fluid Mechanics* 757, 208-230.

Brändle de Motta, J.C., Breugem, W.-P., Gazanion, B, Estivalezes, J.-L., Vincent S. & Climent E. 2013 Numerical modelling of finite-size particle collisions in a viscous fluid. *Physics of Fluids* 25, 083302.

Caiazzo, A. 2008 Analysis of lattice Boltzmann nodes initialisation in moving boundary problems. *Progress in Computational Fluid Dynamics* 8, 3-10.

Chrust, M., Bouchet, G. & Dušek, J. 2013 Numerical simulation of the dynamics of freely falling discs. *Physics of Fluids* 25, 044102.

Clift, R., Grace, J.R. & Weber, M.E. 1978 Bubbles, drops, and particles. *Academic Press*.

D'Humieres, D. 2002 Multiple relaxation time lattice Boltzmann models in three dimensions. *Phil Trans R Soc Lond A* 360, 437-451.

Ern, P., Risso, F., Fabre, D. & Magnaudet, J. 2012 Wake-Induced Oscillatory Paths of Bodies Freely Rising or Falling in Fluids. *Annual Review of Fluid Mechanics* 44, 97-121.

Ern, P. & Brosse, N. 2014 Interaction of two axisymmetric bodies falling side by side at moderate Reynolds numbers. *Journal of Fluid Mechanics* 741, R6.

Feng, J. & Joseph, D.D. 1995 The unsteady motion of solid bodies in creeping flows. *Journal of Fluid Mechanics* 303, 83-102.

Field, S.B., Klaus, M., Moore, M.G. & Nori, F. 1997 Chaotic dynamics of falling disks. *Nature* 388, 252-254.

Fornari, W., Ardekani, M.N. & Brandt, L. 2018 Clustering and increased settling speed of oblate particles at finite Reynolds number. *Journal of Fluid Mechanics* 848, 696-721.

Fortes, A.F., Joseph, D.D. & Lundgren, T.S. 1987 Nonlinear mechanics of fluidization of beds of spherical particles. *Journal of Fluid Mechanics* 177, 467-483.

Guo, Y., Wassgren, C., Ketterhagen, W., Hancock, B. & Curtis, J. 2012a Some computational considerations associated with discrete element modeling of cylindrical particles. *Powder Technol.* 228, 193–198.

Guo, Y., Wassgren, C., Ketterhagen, W., Hancock, B., James, B. & Curtis, J. 2012b A numerical study of granular shear flows of rod-like particles using the Discrete Element Method. *Journal of Fluid Mechanics* 713, 1-26.

Guo, Z., Zheng, C. & Shi, B. 2002 Discrete lattice effects on the forcing term in the lattice Boltzmann method. *Physical Review E* 65, 046308.

He, X. & Luo, L.S. 1997 Lattice Boltzmann Model for the Incompressible Navier–Stokes Equation. *Journal of Statistical Physics* 88, 927-944.

Kodam, M., Bharadwaj, R., Curtis, J., Hancock, B. & Wassgren, C. 2010 Cylindrical object contact detection for use in discrete element method simulations. Part I – Contact detection algorithms. *Chemical Engineering Science* 65, 5852-5862.

Kharrouba, M., Pierson, J.-L. & Magnaudet, J. 2021 Flow structure and loads over inclined cylindrical rodlike particles and fibers. *Physical Review Fluids* 6, 044308.

Ladd, A.J.C. 1994 Numerical simulations of particulate suspensions via a discretized Boltzmann equation. Part 1. Theoretical foundation. *Journal of Fluid Mechanics* 271, 285-309.

Lee, C., Su, Z., Zhong, H., Chen, S., Zhou, M. & Wu, J. 2013 Experimental investigation of freely falling thin disks. Part 2. Transition of three-dimensional motion from zigzag to spiral. *Journal of Fluid Mechanics* 732, 77-104.

Li, M., Zhang, Y., Wang, Y. & Wu, C. 2020 Scaling law of contact time for particles settling in a quiescent fluid. *International Journal of Multiphase Flow* 129, 103317.

Nie, D., Guan, G. & Lin, J. 2021 Interaction between two unequal particles at intermediate Reynolds numbers: A pattern of horizontal oscillatory motion. *Physical Review E* 103, 013105.

Nie, D. & Lin, J. 2020 Simulation of sedimentation of two spheres with different densities in a square tube. *Journal of Fluid Mechanics* 896, A12.

Peng, C., Teng, Y., Hwang, B., Guo, Z.L., Wang, L.-P. 2016 Implementation issues and benchmarking of lattice Boltzmann method for moving rigid particle simulations in a viscous flow. *Computers & Mathematics with Applications* 72, 349-374.

Pierson, J.-L., Hammouti, A., Auguste, F. & Wachs, A. 2019 Inertial flow past a finite-length axisymmetric cylinder of aspect ratio 3: Effect of the yaw angle. *Phsical Review Fluids* 4, 044802.

Tangri, H., Guo, Y. & Curtis, J.S. 2017 Packing of Cylindrical Particles: DEM Simulations and Experimental Measurements. *Powder Technology* 317, 72-82.

Tangri, H., Guo, Y. & Curtis, J.S. 2019 Hopper Discharge of Elongated Particles of Varying Aspect Ratio: Experiments and DEM simulations. *Chemical Engineering Science: X* 4, 100040.

Tchoufag, J., Fabre, D. & Magnaudet, J. 2014 Global linear stability analysis of the wake and path of buoyancy-driven disks and thin cylinders. *Journal of Fluid Mechanics* 740, 278-311.

Tozzi, E.J., Scott, C.T., Vahey, D. & Klingenberg, D.J. 2011 Settling dynamics of asymmetric rigid fibers. *Physics of Fluids* 23, 033301.

Verjus, R., Guillou, S., Ezersky, A. & Angilella, J.-R. 2016 Chaotic sedimentation of particle pairs in a vertical channel at low Reynolds number: Multiple states and routes to chaos. *Physics of Fluids* 28, 123303.

Wang, L.-P., Peng, C., Guo, Z. & Yu, Z. 2016 Lattice Boltzmann simulation of particle-laden turbulent channel flow. *Computers & Fluids*, 226-236.

Wen, B.H., Zhang, C.Y., Tu, Y.S., Wang, C.L. & Fang, H.P. 2014 Galilean invariant fluid-solid

interfacial dynamics in lattice Boltzmann simulations. *Journal of Computational Physics* 266, 161-170.

Willmarth, W.W., Hawk, N.E. & Harvey, R.L. 1964 Steady and Unsteady Motions and Wakes of Freely Falling Disks. *Physics of Fluids* 7, 197-208.

Zastawny, M., Mallouppas, G., Zhao, F. & Wachem, B. 2012 Derivation of drag and lift force and torque coefficients for non-spherical particles in flows. *International Journal of Multiphase Flow* 39, 227-239.

Zhang, Y., Zhang, Y., Pan, G. & Haeri, S. 2018 Numerical study of the particle sedimentation in a viscous fluid using a coupled DEM-IB-CLBM approach. *Journal of Computational Physics* 368, 1-20.

Zhong, H., Chen, S. & Lee, C. 2011 Experimental study of freely falling thin disks: Transition from planar zigzag to spiral. *Physics of Fluids* 23, 110.